\documentclass[]{interact}

\usepackage{epstopdf}% To incorporate .eps illustrations using PDFLaTeX, etc.
\usepackage{subfigure}% Support for small, `sub' figures and tables
\usepackage{todonotes}
\usepackage{xcolor}

\usepackage{natbib}% Citation support using natbib.sty
\bibpunct[, ]{(}{)}{;}{a}{}{,}% Citation support using natbib.sty
\renewcommand\bibfont{\fontsize{10}{12}\selectfont}% Bibliography support using natbib.sty

\theoremstyle{plain}% Theorem-like structures provided by amsthm.sty

\theoremstyle{definition}

\theoremstyle{remark}

\begin{document}

% \articletype{ARTICLE TEMPLATE}

\title{Compressed Sensing for Memory Reduction in Monte Carlo Simulations}

\author{
\name{Ethan Lame\thanks{Corresponding Author - Ethan Lame. Email: lamee@oregonstate.edu}, Camille J. Palmer, Todd Palmer, Ilham Variansyah}
\affil{Nuclear Science and Engineering - Oregon State University, Corvallis, OR}
}

\maketitle

\begin{abstract}
Monte Carlo simulations of neutronic systems are computationally intensive and demand significant memory resources for high-fidelity modeling. Compressed sensing enables accurate reconstruction of signals from significantly fewer samples than traditional methods. The specific implementation of compressed sensing investigated here involves the use of overlapping cells to collect tallies. Increasing the number of samples improves the reconstruction accuracy, although the marginal gains diminish with more samples. Reconstruction quality is strongly influenced by the sparsity parameter used in basis pursuit denoising. Across the three test cases considered, memory reductions of up to 81.25\% (96.25\%) are demonstrated for 2D (3D) reconstructions, with select scenarios achieving reconstruction errors within 1 standard deviation of the corresponding high-fidelity reference results.
\end{abstract}


\section{Introduction}
Monte Carlo neutron transport simulations can be very computationally and temporally expensive to obtain good accuracy. Methods for decreasing the amount of memory needed and/or the computational time are of interest.

Compressed sensing was developed as a method to use far fewer measurements than traditionally necessary to accurately reconstruct a signal \citep{StableSignalRecovery}, and is based on two key principles. The signal to be reconstructed must be sparse in some basis, and there must be incoherence between the measurements and the new basis \citep{compressed_sensing}. A discrete cosine transform works well to represent natural images as sparse, and randomly sampling the signal ensures the incoherence constraint is met \citep{incoherence_constraint}.

Compressed sensing has been previously applied to Monte Carlo neutron transport, with good success \citep{vaquer2016, madsen_2017_disjoint_tallies}. One such approach uses random disjoint tallies, where tally regions are broken into non-contiguous spatial cells that are randomly distributed throughout the problem domain, which reduces memory requirements from covering the whole domain. In contrast, the present work introduces a compressed sensing approach based on overlapping tallies arranged in Cartesian space. Unlike both regular mesh tallies and random disjoint tallies, these overlapping regions allow for localized measurements while still providing sufficient coverage for reconstruction of global flux measurements. The implementation here allows a single basis built from the overlapping regions to cover the vast majority of the problem space. Overlapping coarse cells enables the resolution of small scale details due to the fact that anything contained within the region of overlap contributes to both coarse cells, and its effect is then localized to the small region contained within both overlapping cells. Monte Carlo simulation tallies can quickly grow in number when many variables (space, energy, angle, time) are considered. With compressed sensing, we aim to allow the neutron tracklength flux tallies to be obtained from a simulation without incurring the memory required by a large number of tallies, while ideally maintaining similar standards of statistical uncertainty.

The compressed sensing algorithms are implemented in the Monte Carlo Dynamic Code (MC/DC)~\citep{mcdc}, developed as part of the Center for Exascale Monte Carlo Neutron Transport
(CEMeNT). MC/DC is a performant and scalable software package that is used to develop and test novel algorithms for neutron transport applications. Written in Python, MC/DC also uses the Numba compiler to accelerate performance and has the capability to run on CPUs and GPUs, with support for parallelism via the use of \texttt{mpi4py} \citep{mpi4py}.

\subsection{Compressed Sensing}
Given a signal $\mathbf{f}$, a transformation can be applied to $\mathbf{f}$ to obtain a sparse signal $\mathbf{x}$ - the representation of $\mathbf{f}$ in a sparse basis. We can also randomly sample $\mathbf{f}$ to obtain measurements of the signal, $\mathbf{b}$. Assuming that the transformation operation was performed with the application of some matrix $\mathbf{T}$, and the sampling performed with some matrix $\mathbf{S}$ we can write the following:
\begin{align}
    \bm{Sf} &= \bm{b}\\
    \bm{S T^{-1} T f} &=\bm{b}\\
    \bm{Tf}&=\bm{x}\\
    \bm{A}&= \bm{ST^{-1}}\label{eq:A=ST-1}\\
    \bm{Ax} &= \bm{b} \label{eq:Ax=b}
\end{align}

The sensing matrix $\mathbf{A}$ encodes the sampling and inverse transformation, while $\mathbf{x}$ is the sparse representation of the signal of interest. We use this final equation to search for a sparse signal $\mathbf{x}$, such that when $\mathbf{x}$ is inversely transformed and sampled (i.e. when $\mathbf{A}$ is applied), the result matches the measurements $\mathbf{b}$. The only information about the problem that is known prior to this search is the basis in which the search will take place, the sampling method $\mathbf{S}$, and the measurements of the signal, $\mathbf{b}$.

\subsubsection{Optimization}
In this work, we choose the transformation $\mathbf{T}$ to be the discrete cosine transform (DCT), and $\mathbf{T^{-1}}$ the inverse discrete cosine transform (IDCT). Natural images are generally sparse in the DCT basis \citep{incoherence_constraint}, which is important for the compressed sensing methods to work well.

To use Equation \ref{eq:Ax=b} to search in the DCT domain, we need to apply some optimization constraints to the problem. There are many optimizations that could be used in order to search in the domain for a vector that matches the measurements of the signal. Compressed sensing may use basis pursuit \citep{compressed_sensing}, in which a sparse vector is chosen to minimize the difference between the chosen vector and the measurements. A modification to the basis pursuit optimization is also often used, called basis pursuit denoising \citep{basis_pursuit_denoising}. This is especially useful when the measurements may be noisy, as it allows for a balance between matching the noisy measurements and a measure of the sparsity of the signal. We choose to use basis pursuit denoising, which takes the following form:

\begin{equation}
    {\displaystyle \min _{x}\left({\frac {1}{2}}\|Ax-b\|_{2}^{2}+\lambda \|x\|_{1}\right).} \label{eq:basis pursuit denoising}
\end{equation}

The first term in the optimization constraint is the accuracy term. If $\mathbf{Ax} = \mathbf{b}$, then the vector $\mathbf{x}$ exactly matches the sparse representation of $\mathbf{f}$, so the optimization may select that value of $\mathbf{x}$, depending on the weight of the second term. The second term controls the importance of sparsity of $\mathbf{x}$. With Monte Carlo simulations, we often don't want to exactly match the noisy measurements. Instead, we note that the true signal is sparse in this basis, so we also emphasize the sparsity of $\mathbf{x}$. A higher value of $\lambda$ prioritizes sparsity over match to measurements, and a lower value of $\lambda$ prioritizes matching the measurements over sparsity.

\section{Methods}
\subsection{Constructing the Matrices}
\subsubsection{Sampling Matrix $\mathbf{S}$}
The sampling matrix $\mathbf{S}$ generates random samples of the signal of interest. The most common tally structures in use in Monte Carlo codes are high resolution structured Cartesian meshes \citep{serpent,mcnp,openmc}. We instead use overlapping cells to store the tallies. This allows coverage of the entire problem with fewer cells than the high resolution Cartesian mesh, reducing the memory cost associated with the tallies, while still capturing small-scale characteristics of the problem via the small regions that are contained in multiple cells.

The center points of the coarse cells are placed upon the problem space by a simple Halton sequence, a low-discrepancy pseudorandom method to ensure that the problem space is more evenly covered than by other ``true'' random sampling techniques \citep{Halton}. The size of the cells is chosen before the problem is executed. Advisable constraints for the size are such that the entire problem space may be covered by the chosen number of cells, while ensuring that the cells are small enough to resolve fine detail. Enforcing that the entire problem space is covered may also be a good constraint, but this was not done in this work.

\begin{figure}[h!]
    \centering
    \includegraphics[width=0.45\linewidth]{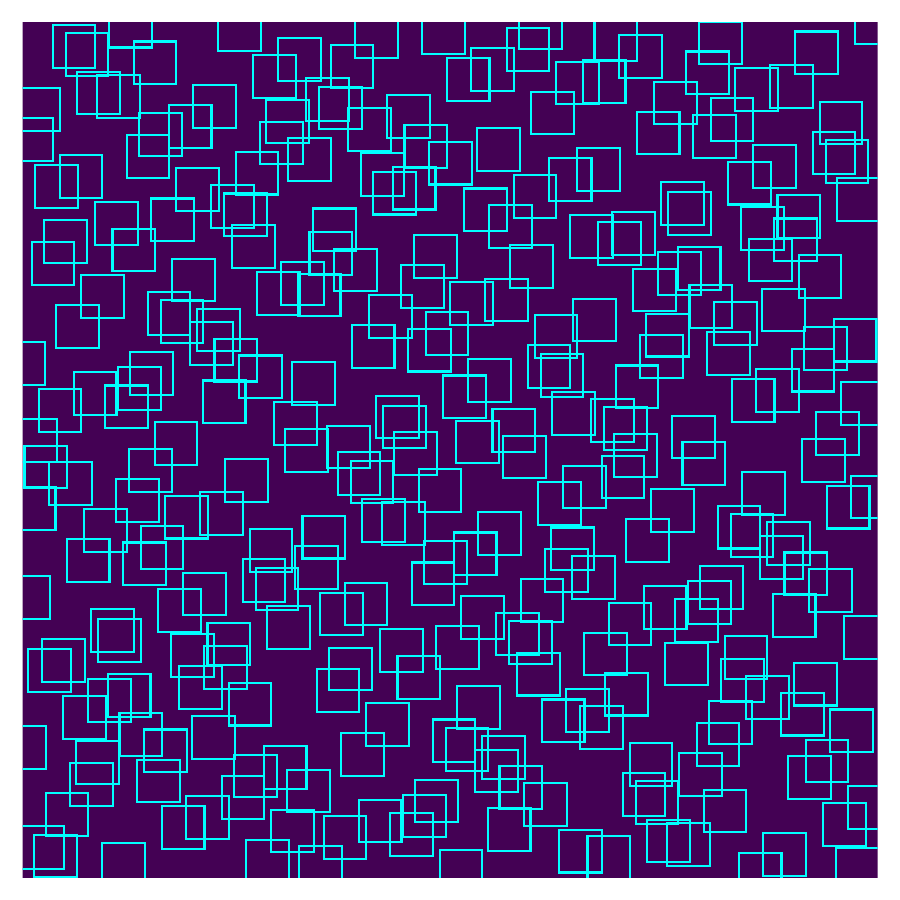}
    \includegraphics[width=0.45\linewidth]{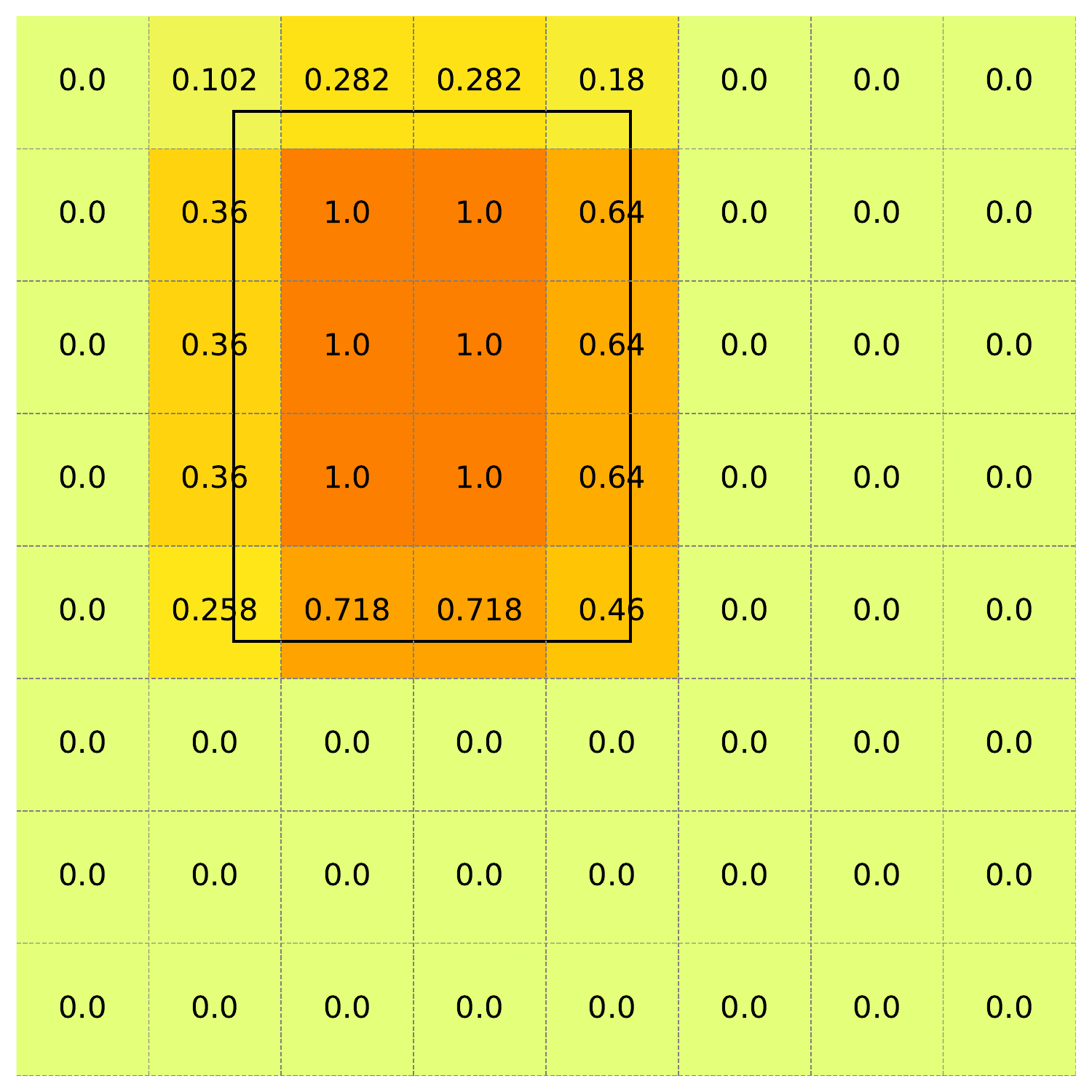}
    \caption{(Left) An example of the overlapping cells in 2D, covering the vast majority of the problem space. (Right) A single coarse bin overlaid on a Cartesian mesh, with redistribution values shown for each cell. The matrix of redistribution values would be flattened and would be one row of the sampling matrix $\mathbf{S}$.}
    \label{fig:overlapping cells}
\end{figure}

An illustrative example of these overlapping cells is shown in Figure \ref{fig:overlapping cells}. The sampling matrix $\mathbf{S}$ is constructed as follows. The coarse overlapping bins are overlaid on a Cartesian mesh of some size. For each coarse bin, the fraction of each standard Cartesian cell's area that lies within the coarse bin is recorded, and all cell values are recorded as a row vector, with as many elements as there are standard cells. This is repeated for each coarse bin, and the rows are stacked on top of each other to form the sampling matrix such that it has as many rows as there are coarse bins and as many columns as standard cells. When post-processing data for compressed sensing, multiplying this sampling matrix by a flattened version of some signal effectively redistributes its data into the coarse bins. For an \textit{in situ} implementation of compressed sensing, only the sampling matrix is obtained from this process and is used in the construction of the sensing matrix, described later.

\subsubsection{Measurement Vector $\mathbf{b}$}
The measurements of the system are collected as tracklength tallies for each coarse bin. As particles stream through space, they trace out their paths within one or more coarse bins. A bin’s tally score increases in proportion to the length of the particle’s path that lies within that bin. A single particle can tally into multiple bins if its path crosses a region of overlap between the bins.

\subsubsection{Sensing Matrix $\mathbf{A}$}
The sensing matrix $\mathbf{A}$ encodes both the sampling of and  transformation from the DCT domain. We established in Equation \ref{eq:A=ST-1} that $\mathbf{A}$ can be constructed from the product of two matrices, with $\mathbf{T^{-1}}$ being the inverse discrete cosine transform (IDCT). For higher resolution reconstructions, the construction of the square matrix $\mathbf{T^{-1}}$ grows prohibitively large.
Instead of explicitly forming and multiplying $\mathbf{S}$ by $\mathbf{T^{-1}}$, we apply the corresponding discrete cosine transform directly using scipy's \texttt{spfft.dctn} function, which performs the same operation efficiently without constructing the matrix.

\begin{equation}
    \mathbf{A = ST^{-1}} \Rightarrow \mathbf{A}^T = (\mathbf{ST^{-1}})^T = (\mathbf{T^{-1}})^T \mathbf{S}^T = \mathbf{TS}^T
\end{equation}
\begin{equation}
    \mathbf{A} = (\mathbf{TS}^T)^T, \label{eq:transpose magic}
\end{equation}
where superscript $T$ indicates transpose. The IDCT and DCT matrices are orthogonal, so $\mathbf{(T^{-1})}^T = \mathbf{T}$. In equation \ref{eq:transpose magic}, the multiplication by the transform matrix $\mathbf{T}$ is instead done by applying \texttt{spfft.dctn} to the matrix $\mathbf{S}$.

\subsection{Reconstruction}
The measurement vector $\mathbf{b}$ is obtained from the simulation as the particles tally into the coarse bins. Then, a desired resolution for the reconstruction is chosen and the sensing matrix is constructed. The basis pursuit denoising optimization is performed with the CVXPY package for Python \citep{CVXPY}, and the sparse vector $\mathbf{x}$ is obtained, which is then inversely transformed to yield the signal $\mathbf{f}$.

The reconstruction error is quantified as the relative $\ell^2$ error between the reconstruction ($\mathbf{r}$) and the reference solution ($\mathbf{f}$):

\begin{equation}
    {\rm error} = \frac{\|\mathbf{f - r}\|_{2}}{\|\mathbf{f}\|_{2}}.
\end{equation}

This reconstruction error is compared with the global standard deviation of the reference solution as a means to quantify the accuracy of the reconstruction.

\section{Results}
\begin{figure}[h!]
    \centering
    \includegraphics[width=0.7\linewidth]{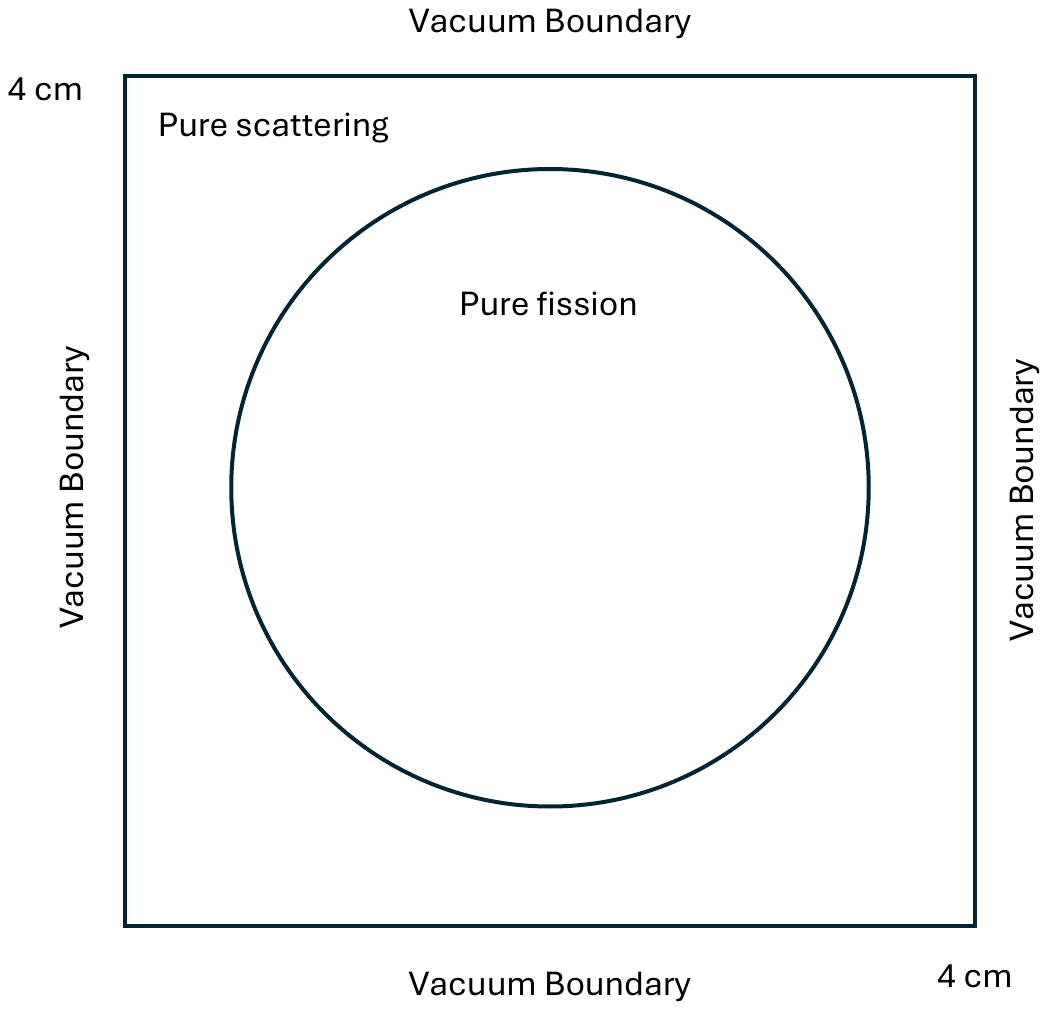}
    \caption{Geometry of the pure-fission sphere problem. The sphere has a radius of 1.5 cm and its center is at the center of the cube with a side length of 4 cm.}
    \label{fig:sphere geometry}
\end{figure}

Three test problems were simulated with MC/DC to evaluate the performance of reconstructions from basis pursuit denoising. The first problem is a very simple pure-fission sphere inside a pure-scattering cube. This simple geometry allows for easy testing and troubleshooting of the code. The geometry and material properties are presented in Figure \ref{fig:sphere geometry} and Table \ref{tab:sphere properties}. This problem uses a default mesh resolution of 40$\times$40 cells in 2D and 20$\times$20$\times$20 cells in 3D. The cells are uniform, and cover a space of 4$\times$4 cm in 2D and 4$\times$4$\times$4 cm in 3D. We quantify the memory reduction by comparing the number of coarse bins used to the number of cells used to compute the reference solution.

\begin{figure}[h!]
    \centering
    \includegraphics[width=0.5\linewidth]{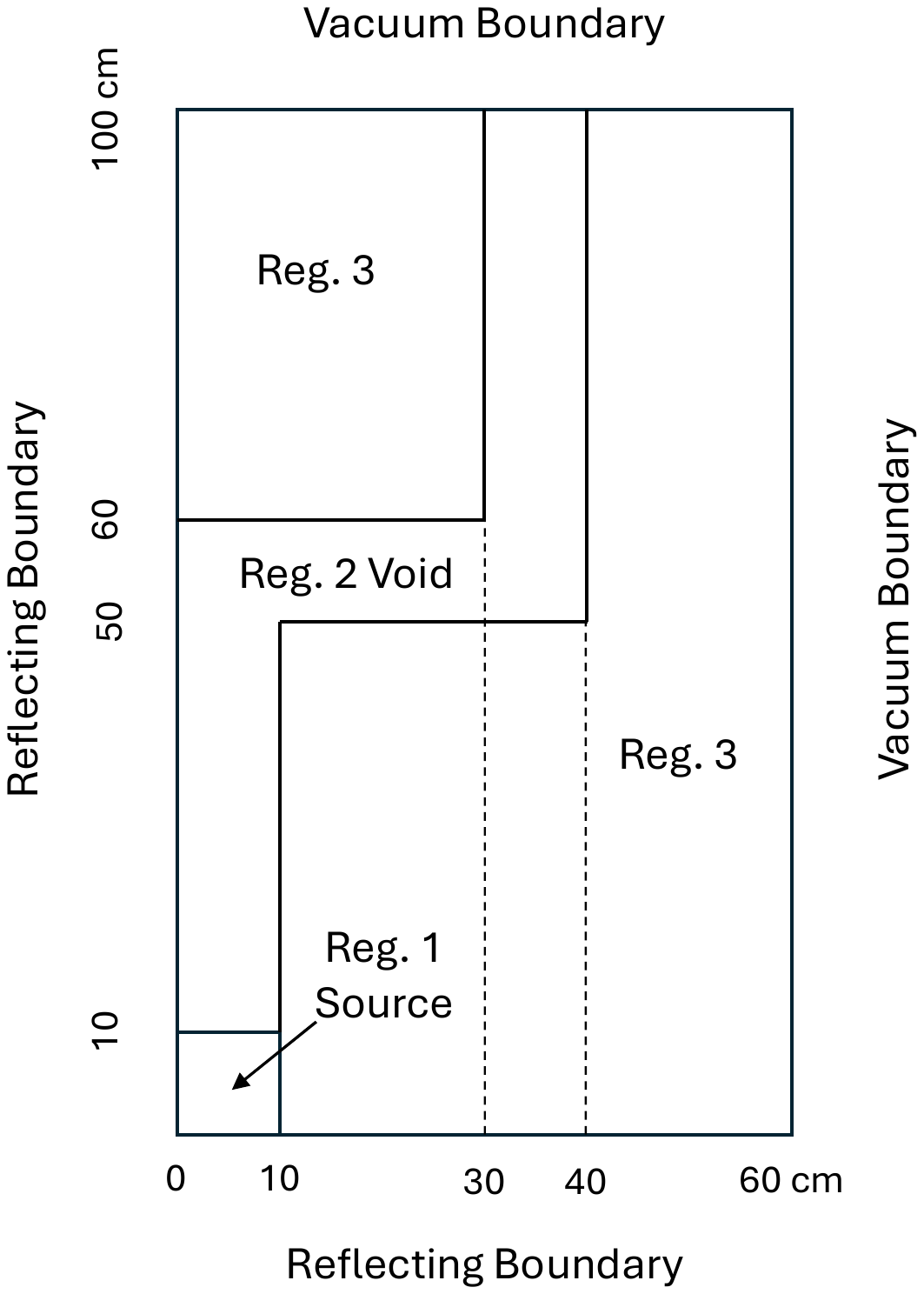}
    \caption{The geometry in the x-y plane of the Kobayashi dog-leg benchmark.}
    \label{fig:kobayashi geometry}
\end{figure}

The second problem is based on the Kobayashi dog-leg benchmark problem \citep{Kobayashi}. The geometry and material properties are shown in Figure \ref{fig:kobayashi geometry} and Table \ref{tab:kobayashi properties}. This problem uses a default mesh resolution of 30$\times$50 cells in 2D and 15$\times$25$\times$15 cells in 3D. The cells are uniform, and cover a space of 60$\times$100 cm in 2D and 60$\times$100$\times$60 cm in 3D.

\begin{figure}[h!]
    \centering
    \includegraphics[width=0.7\linewidth]{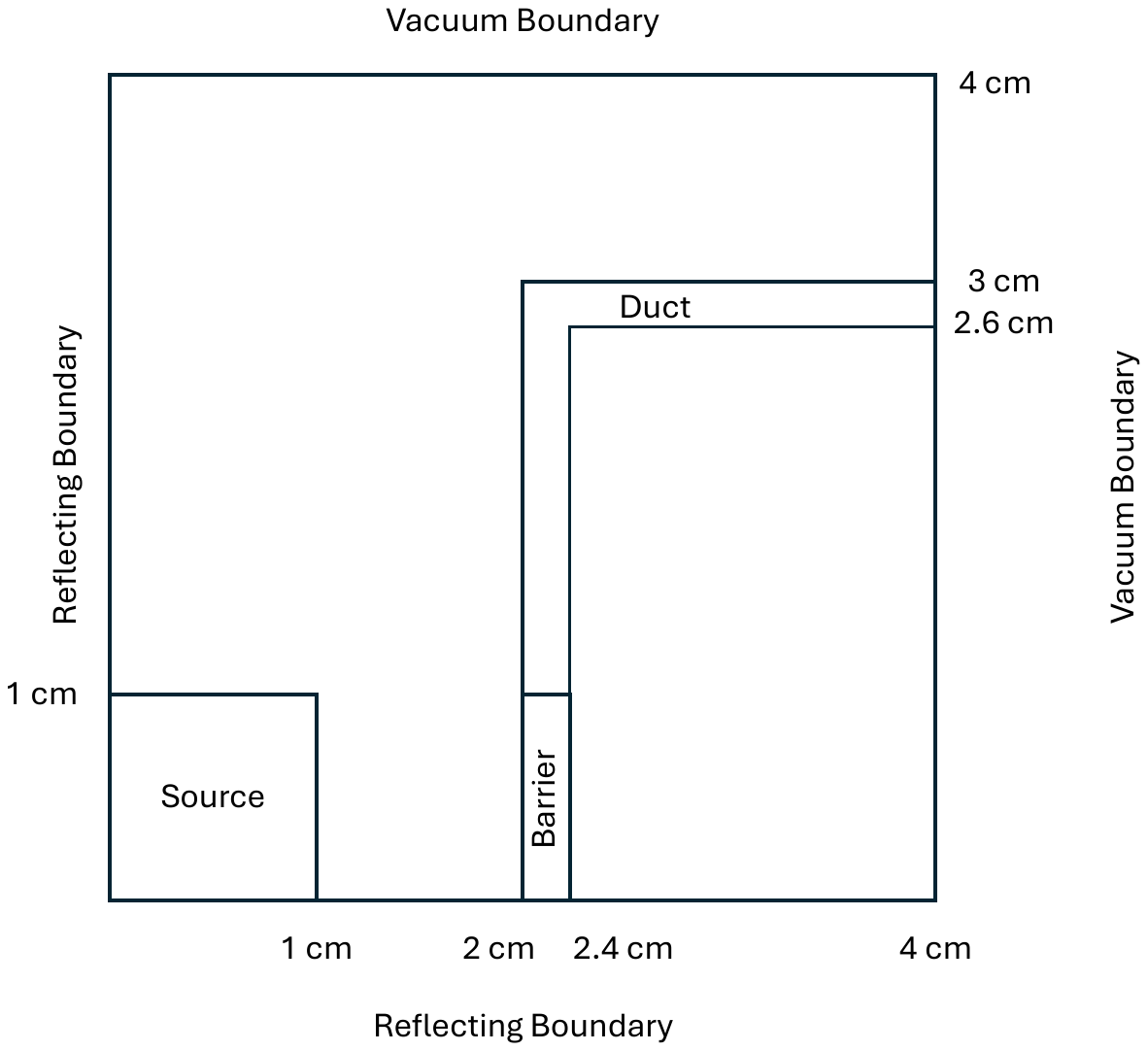}
    \caption{The geometry of the modified Cooper-Larsen problem. The barrier extends 1 cm high, in line with the edge of the source.}
    \label{fig:cooper larsen geometry}
\end{figure}

The third problem is a combination of the problems presented in \cite{Cooper_Larsen}. It contains a highly scattering barrier region and a void duct region, allowing us to see how well the compressed sensing may reconstruct areas that receive little to no particle flux. The geometry and material properties are shown in Figure \ref{fig:cooper larsen geometry} and Table \ref{tab:cooper larsen properties}. This problem uses a default mesh resolution of 40$\times$40 cells, and is only in 2D. The cells are uniform, and cover a space of 4$\times$4 cm. All three problems use $10^4$ particle histories for each simulation.

\begin{table}[h!]
    \centering
    \begin{tabular}{|c|c|c|c|c|}
    \hline
        Region & $\Sigma_t$ & $\Sigma_s$ & $\Sigma_f$ & $\nu$ \\ \hline\hline
        Sphere & 1.0 & 0 & 1.0 & 1.1 \\ \hline
        Outside of Sphere & 1.0 & 1.0 & 0 & N/A \\ \hline
    \end{tabular}
    \caption{Material properties of the pure-fission sphere problem. Cross sections are in units of cm$^{-1}$. Particles are spawned uniformly throughout the problem.}
    \label{tab:sphere properties}
\end{table}

\begin{table}[h!]
    \centering
    \begin{tabular}{|c|c|c|c|}
    \hline
        Region & Source (n cm$^{-3}$ s$^{-1}$) & $\Sigma_t$ (cm$^{-1}$) & $c = \Sigma_s / \Sigma_t$ \\ \hline\hline
        1 & 1 & 0.1 & 0.5 \\ \hline
        2 & 0 & $10^{-4}$ & 0.5 \\ \hline
        3 & 0 & 0.1 & 0.5 \\ \hline
    \end{tabular}
    \caption{Material properties of the Kobayashi problem. Cross sections are in units of cm$^{-1}$. Particles are spawned uniformly throughout the problem.}
    \label{tab:kobayashi properties}
\end{table}

\begin{table}[h!]
    \centering
    \begin{tabular}{|c|c|c|}
    \hline
        Region & $\Sigma_t$ & $c = \Sigma_s / \Sigma_t$ \\ \hline\hline
        Barrier & 5 & 0.8 \\ \hline
        Duct & 0.01 & 0.8 \\ \hline
        Elsewhere & 1 & 0.8 \\ \hline
    \end{tabular}
    \caption{Material properties of the modified Cooper-Larsen problem. Cross sections are in units of cm$^{-1}$. Particles are spawned uniformly throughout the problem.}
    \label{tab:cooper larsen properties}
\end{table}

\subsection{Sphere Reconstructions Using Different Values of $\lambda$}
\subsubsection{2D Results}
\begin{figure}[h]
    \centering
    \includegraphics[width=0.72\linewidth]{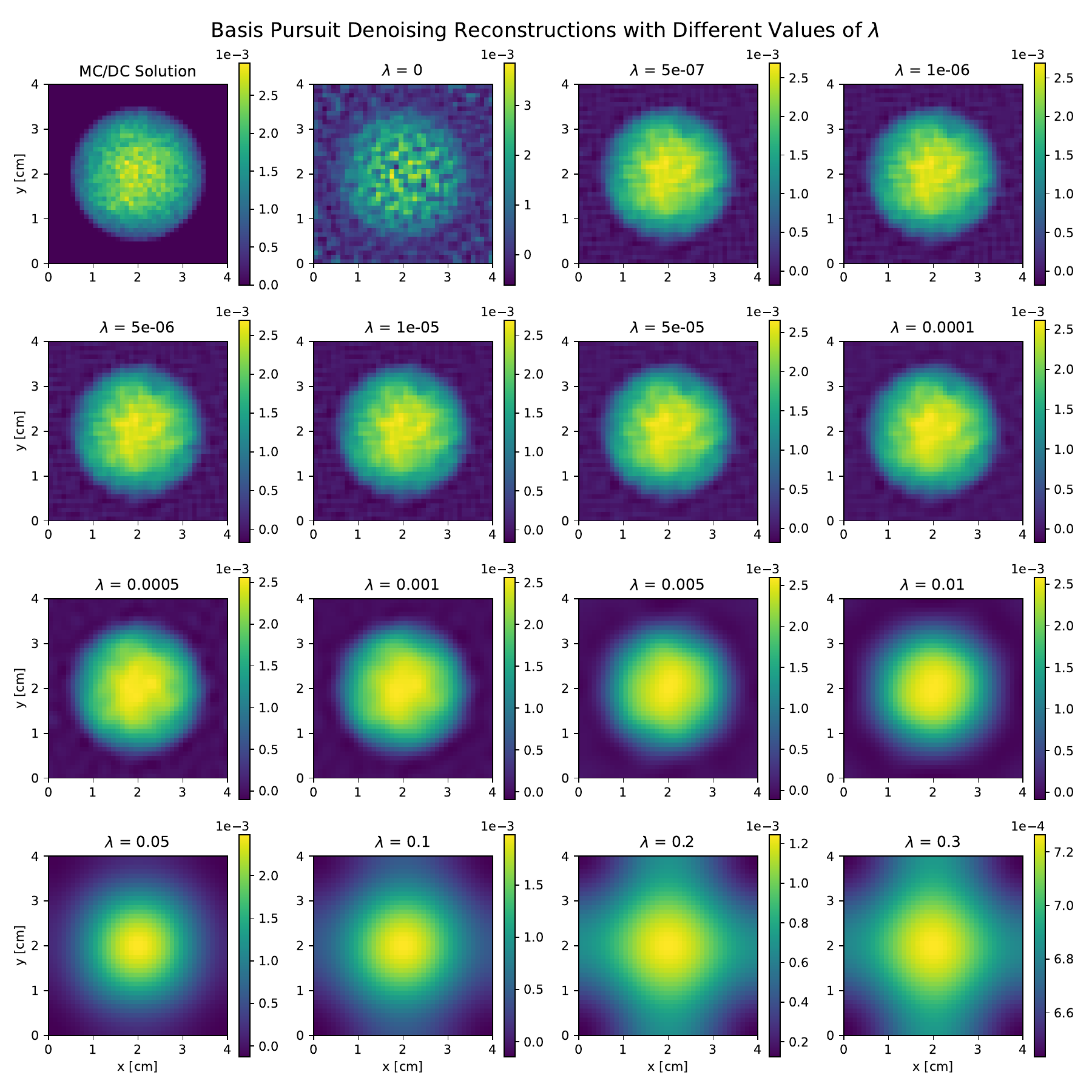}
    \includegraphics[width=0.72\linewidth]{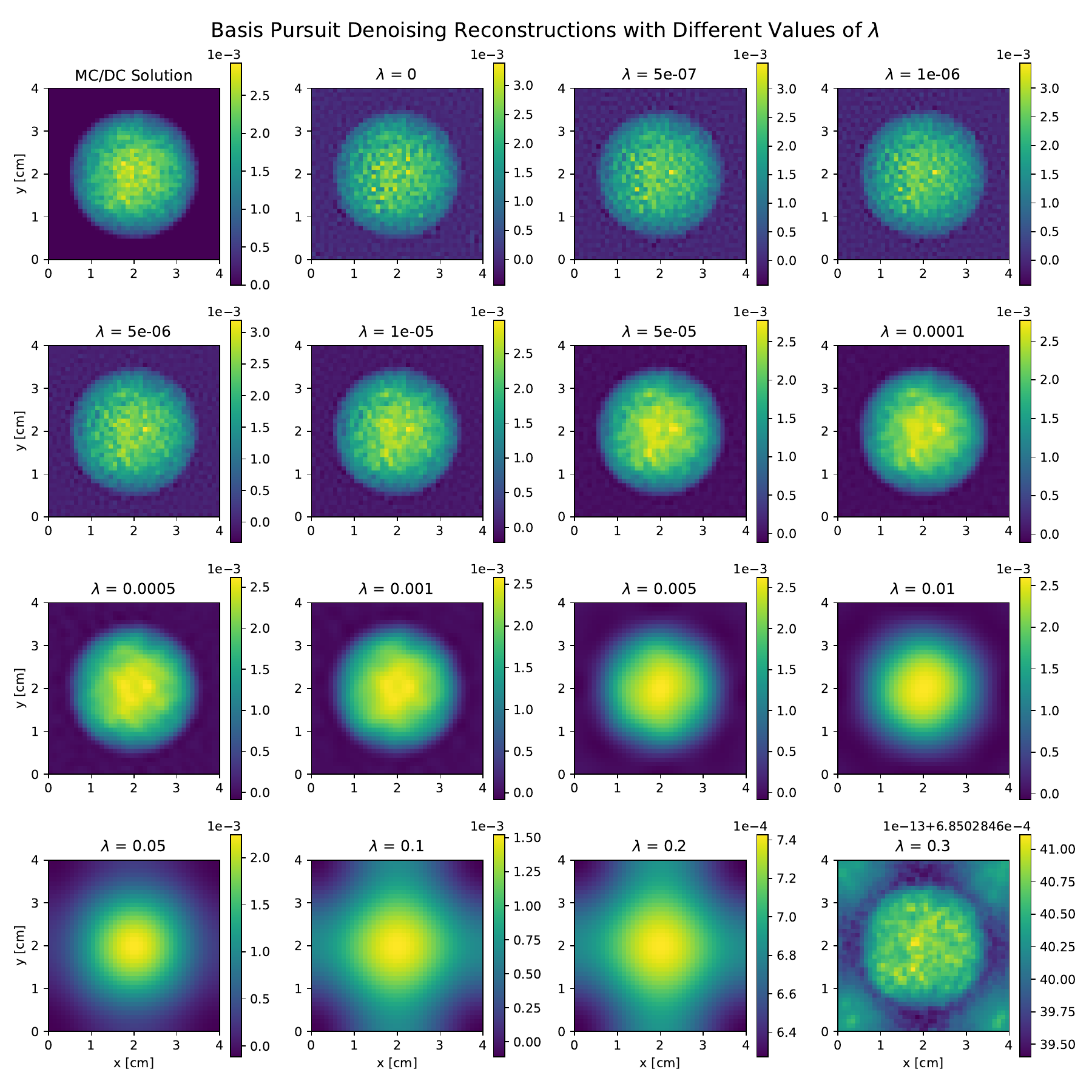}
    \caption{Reconstructions of the sphere problem in 2D using basis pursuit denoising for different values of $\lambda$. 300 bins of size 3$\times$3 pixels on the left; 1000 bins of size 2$\times$2 pixels on the right. The reference solution is shown in the top left of each set, run with $10^4$ particles. Errors associated with each reconstruction are plotted in Figure \ref{fig:2D sphere errors}.}
    \label{fig:2D sphere recons}
\end{figure}

\begin{figure}[h!]
    \centering
    \includegraphics[width=0.65\linewidth]{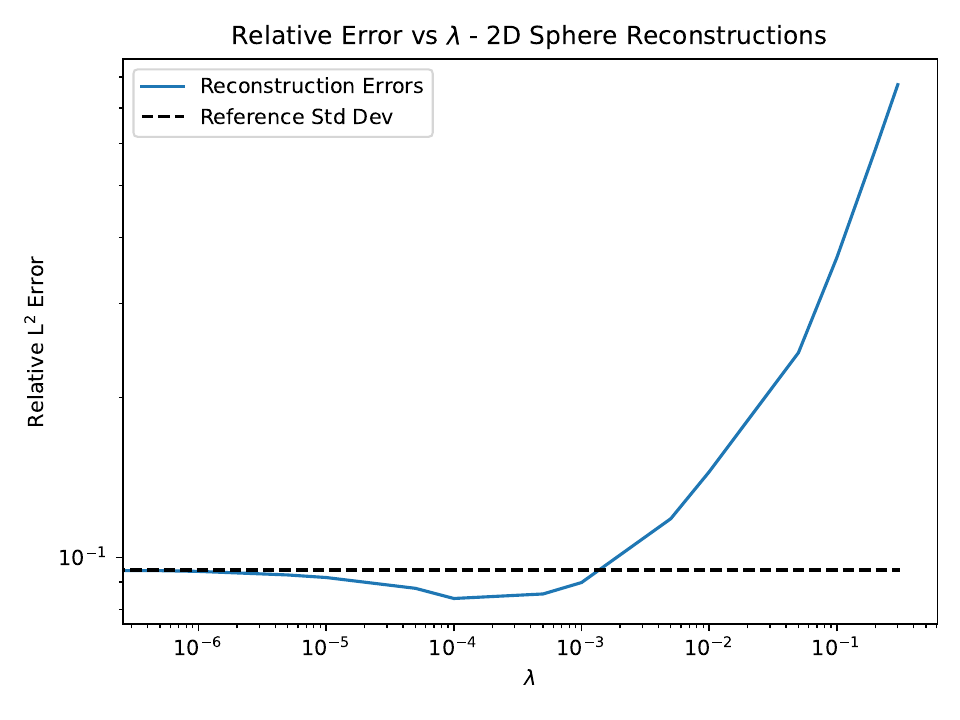}
    \includegraphics[width=0.65\linewidth]{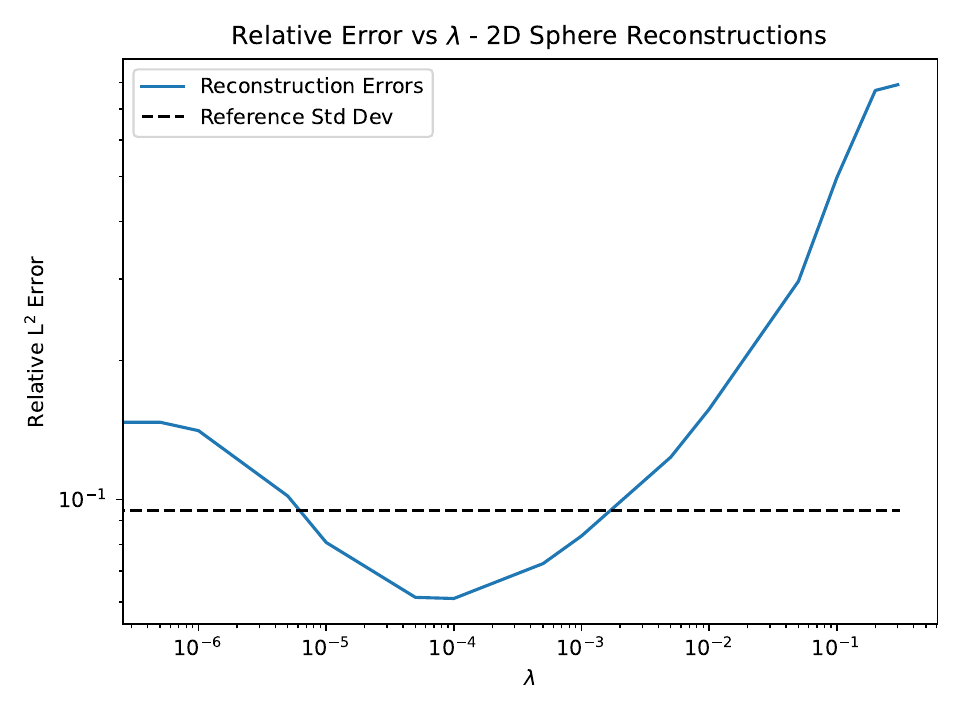}
    \caption{The errors associated with each reconstruction of the 2D sphere in Figure \ref{fig:2D sphere recons}. Top: 300 3$\times$3 bins, 81.25\% memory reduction. Bottom: 1000 2$\times$2 bins, 37.5\% memory reduction.}
    \label{fig:2D sphere errors}
\end{figure}

The reconstructions for the 2D sphere problem are shown in Figure \ref{fig:2D sphere recons}, with statistical relative errors in Figure \ref{fig:2D sphere errors}. At large values of $\lambda$, we see symmetry in the reconstructions along both the x and y axes, but for other values of $\lambda$, the reconstructions appear to be radially symmetric, matching the reference solution. It appears that the main difference between values of $\lambda$ comes from the sharpness of the edge of the sphere and the magnitude of the background flux. Small $\lambda$ have a nonzero background flux, and large $\lambda$ have a fuzzy border between the sphere and the background - somewhere in the middle is the best choice to satisfy the basis pursuit denoising. The values of $\lambda$ were chosen to span a wide range, with basis pursuit denoising typically using a value of 0.5. A few preliminary tests using $\lambda = 0.5$ were not promising, so smaller values were chosen for these reconstructions.

\newpage
\subsubsection{3D Results}
\begin{figure}
    \centering
    \includegraphics[width=\linewidth]{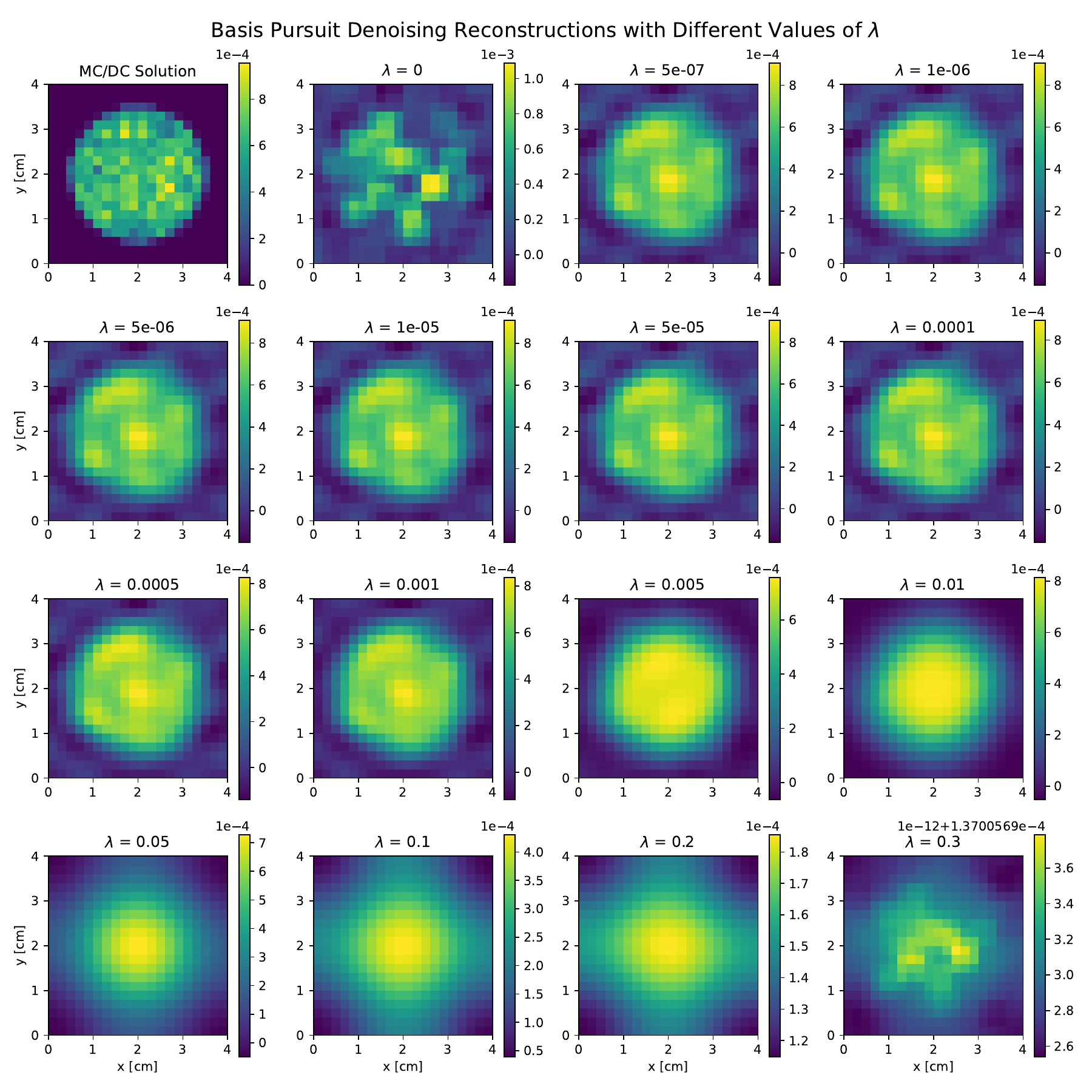}    \caption{Slices of 3D reconstructions of the sphere using basis pursuit denoising. Results with 300 bins. All bins are of size 3$\times$3$\times$3 pixels. Errors associated with these reconstructions are plotted in Figure \ref{fig:3D sphere errors}.}
    \label{fig:3D sphere recons 300}
\end{figure}

\begin{figure}
    \centering
    \includegraphics[width=\linewidth]{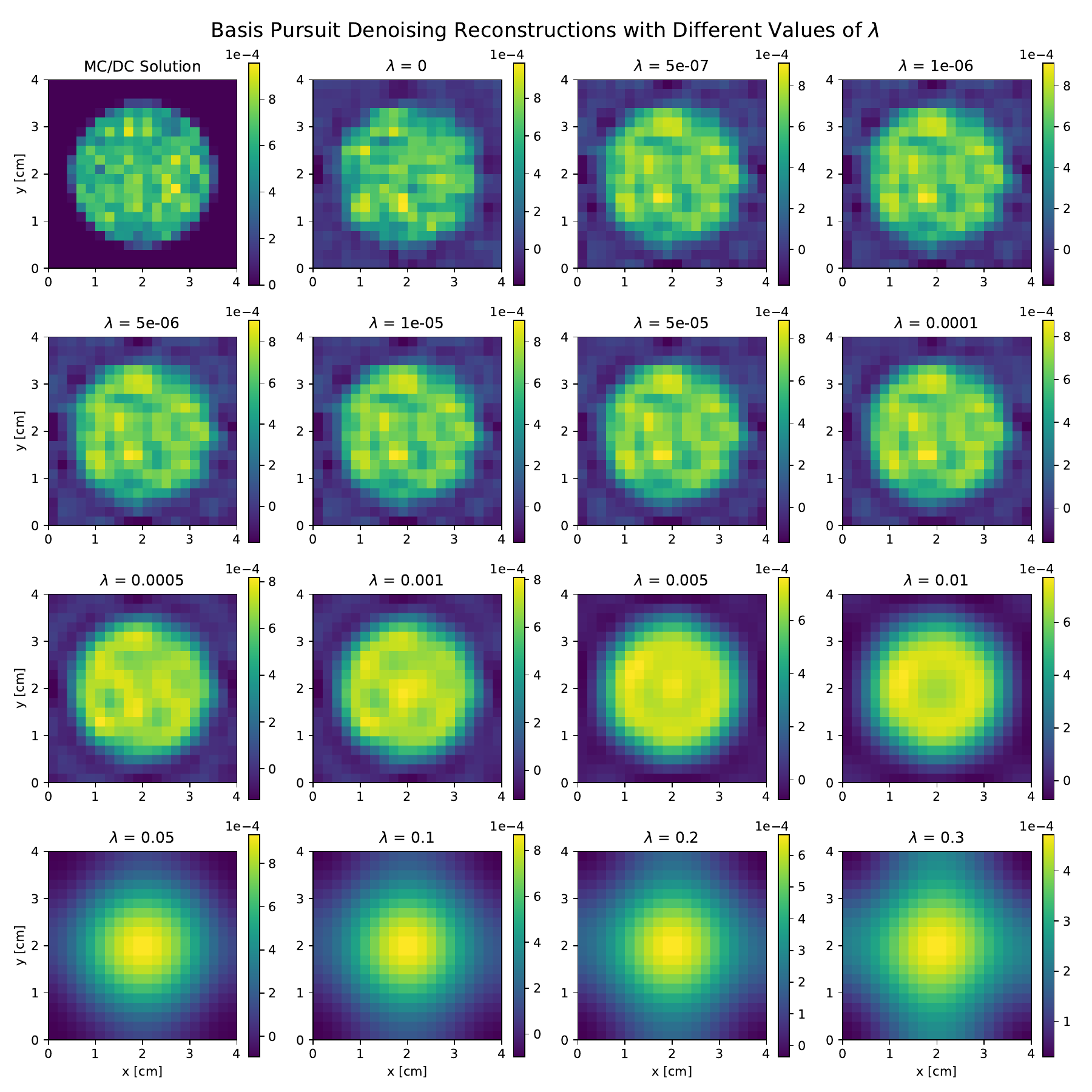}    \caption{Slices of 3D reconstructions of the sphere using basis pursuit denoising. Results with 1000 bins. All bins are of size 3$\times$3$\times$3 pixels. Errors associated with these reconstructions are plotted in Figure \ref{fig:3D sphere errors}.}
    \label{fig:3D sphere recons 1000}
\end{figure}

\begin{figure}
    \centering
    \includegraphics[width=\linewidth]{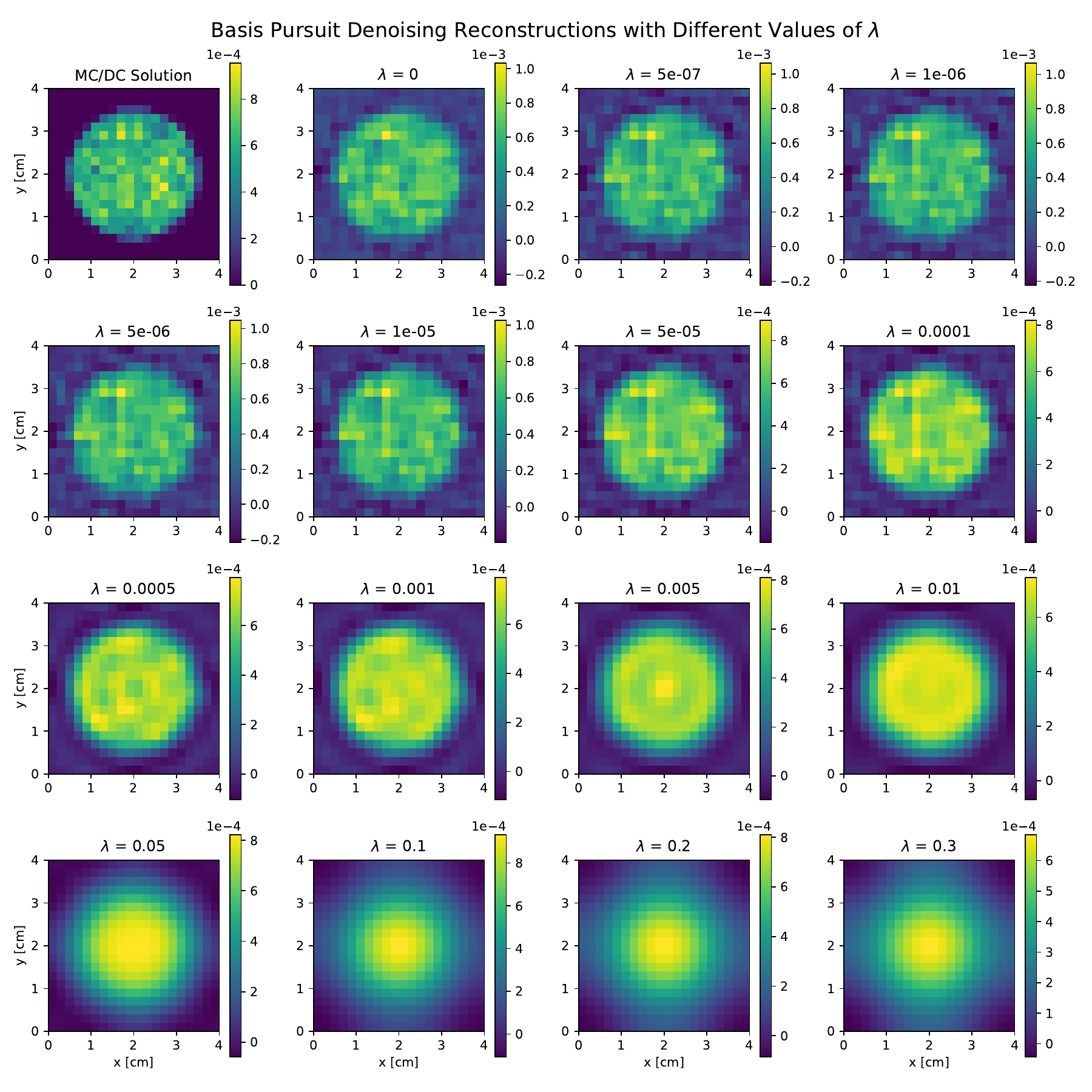}    \caption{Slices of 3D reconstructions of the sphere using basis pursuit denoising. Results with 1600 bins. All bins are of size 3$\times$3$\times$3 pixels. Errors associated with these reconstructions are plotted in Figure \ref{fig:3D sphere errors}.}
    \label{fig:3D sphere recons 1600}
\end{figure}

\begin{figure}[h!]
    \centering
    \includegraphics[width=0.65\linewidth]{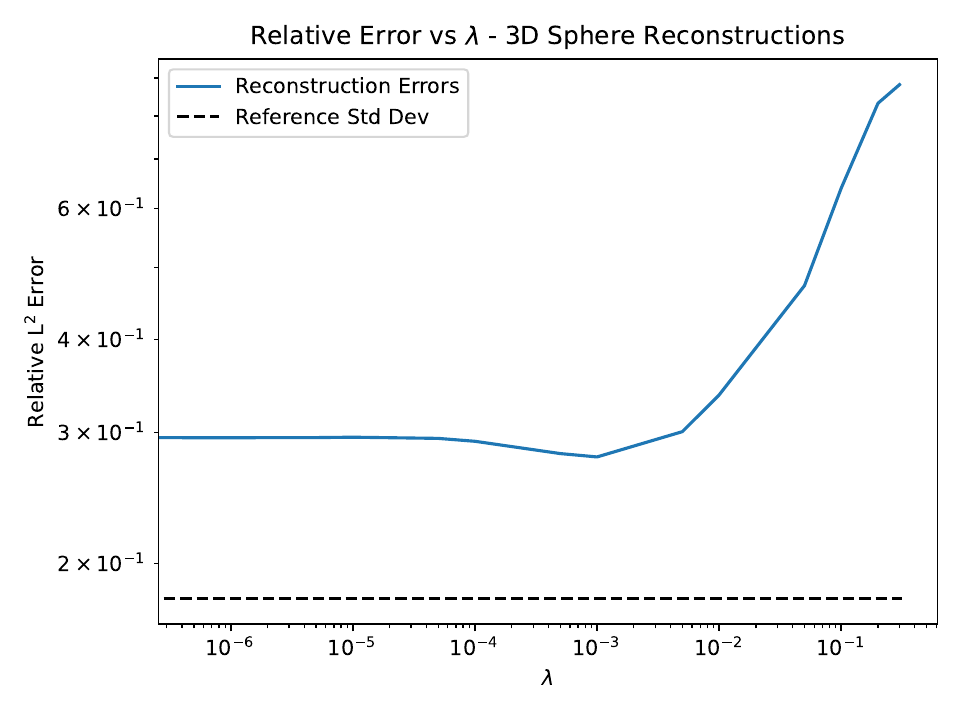}
    \includegraphics[width=0.65\linewidth]{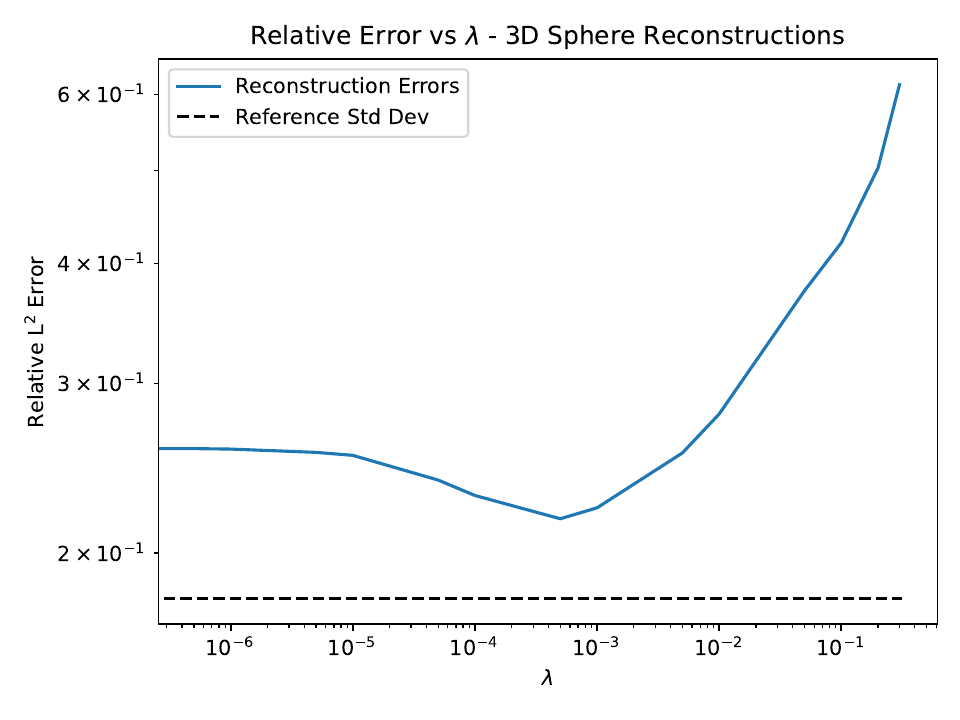}
    \includegraphics[width=0.65\linewidth]{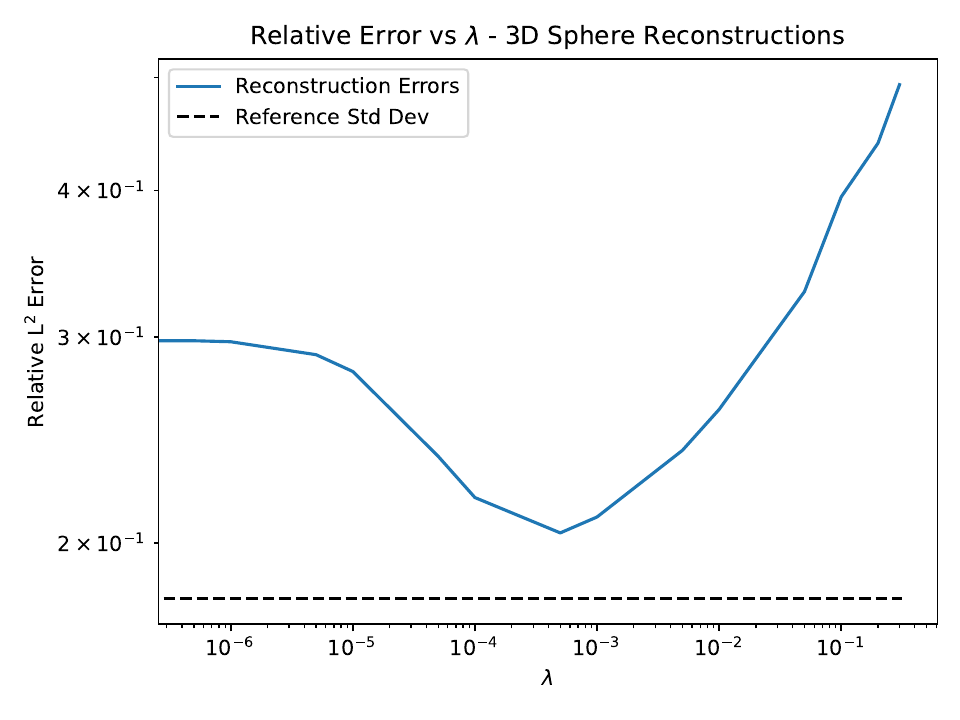}
    \caption{Errors associated with reconstructions of the sphere shown in Figures \ref{fig:3D sphere recons 300}, \ref{fig:3D sphere recons 1000}, and \ref{fig:3D sphere recons 1600}. Top: 300 3$\times$3$\times$3 bins, 96.25\% memory reduction. Middle: 1000 3$\times$3$\times$3 bins, 87.5\% memory reduction. Bottom: 1600 3$\times$3$\times$3 bins, 80\% memory reduction.}
    \label{fig:3D sphere errors}
\end{figure}

The reconstructions for the 3D sphere problem are shown in Figures \ref{fig:3D sphere recons 300}, \ref{fig:3D sphere recons 1000}, and \ref{fig:3D sphere recons 1600}, with statistical relative errors in Figure \ref{fig:3D sphere errors}.


\subsection{Kobayashi Problem Reconstructions Using Different Values of $\lambda$}
\subsubsection{Statistical Relative Errors in 2D Reconstructions}
\begin{figure}[h!]
    \centering
    \includegraphics[width=0.65\linewidth]{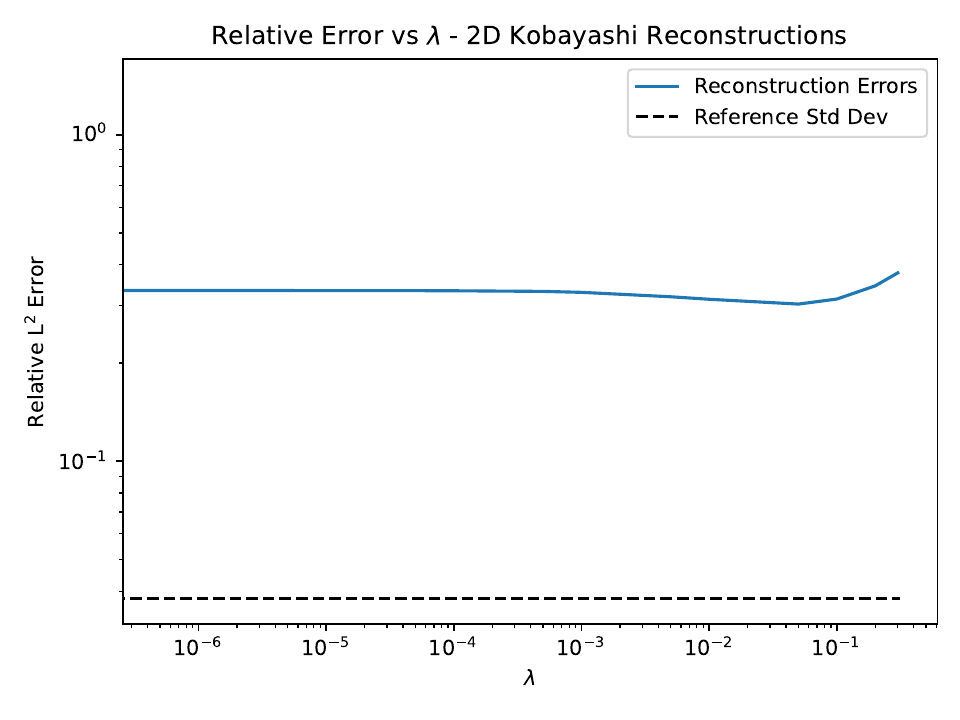}
    \includegraphics[width=0.67\linewidth]{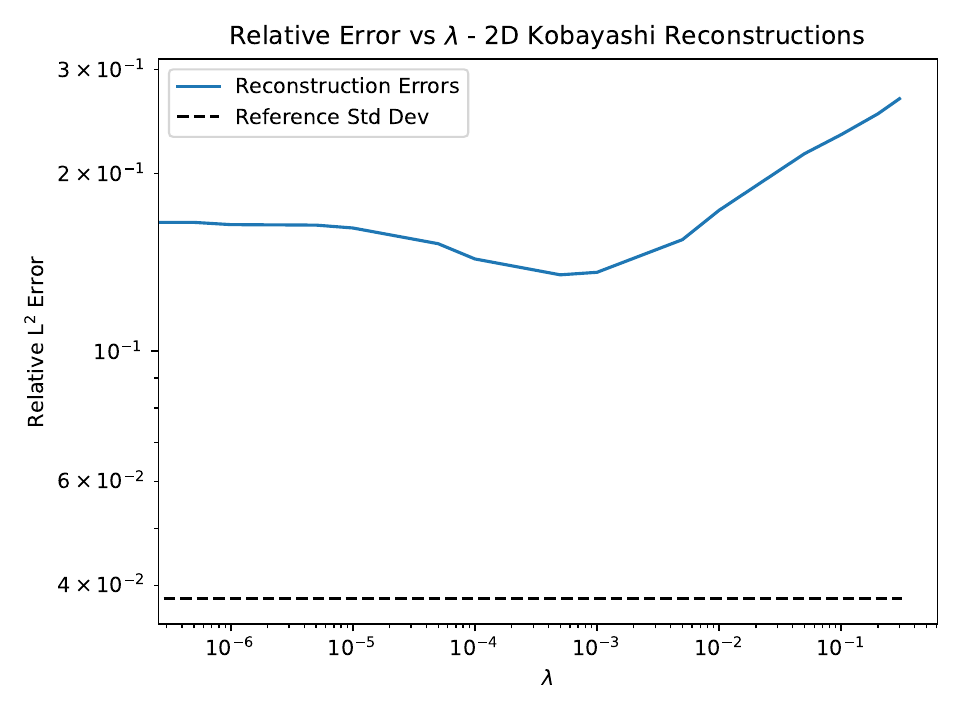}
    \includegraphics[width=0.65\linewidth]{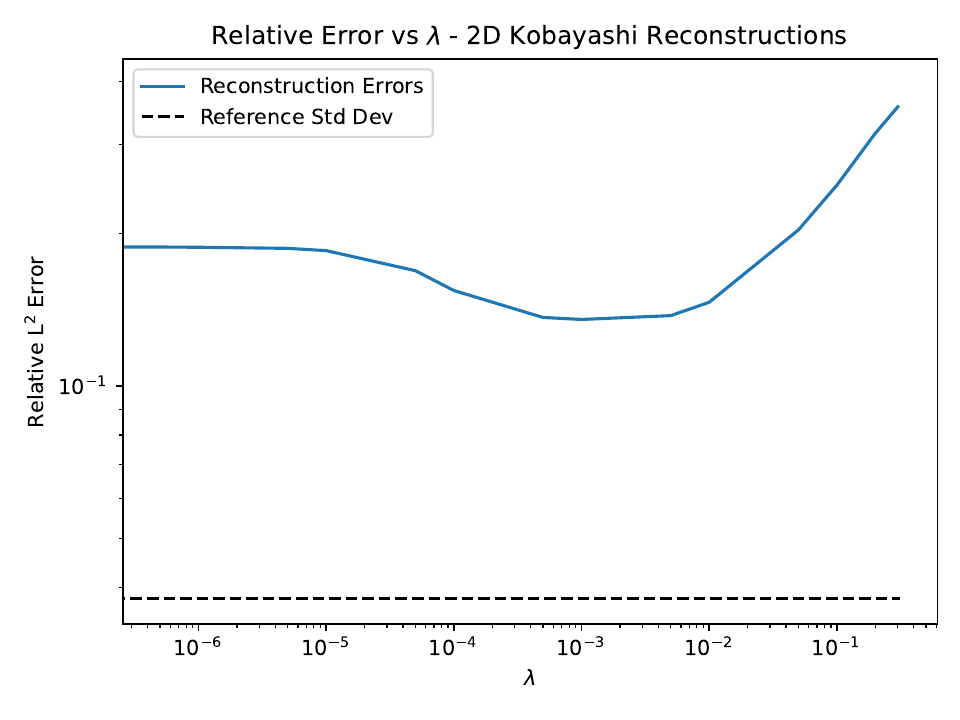}
    \caption{Errors associated with reconstructions of the Kobayashi problem. Top: 300 3$\times$3 bins, 80\% memory reduction. Middle: 1000 3$\times$3 bins, 33.3\% memory reduction. Bottom: 1000 2$\times$2 bins, 33.3\% memory reduction.}
    \label{fig:2D kobayashi errors}
\end{figure}

The statistical errors associated with reconstructions of the Kobayashi problem are shown in Figure \ref{fig:2D kobayashi errors}.

\subsubsection{Statistical Relative Errors in 3D Reconstructions}
\begin{figure}[h!]
    \centering
    \includegraphics[width=0.65\linewidth]{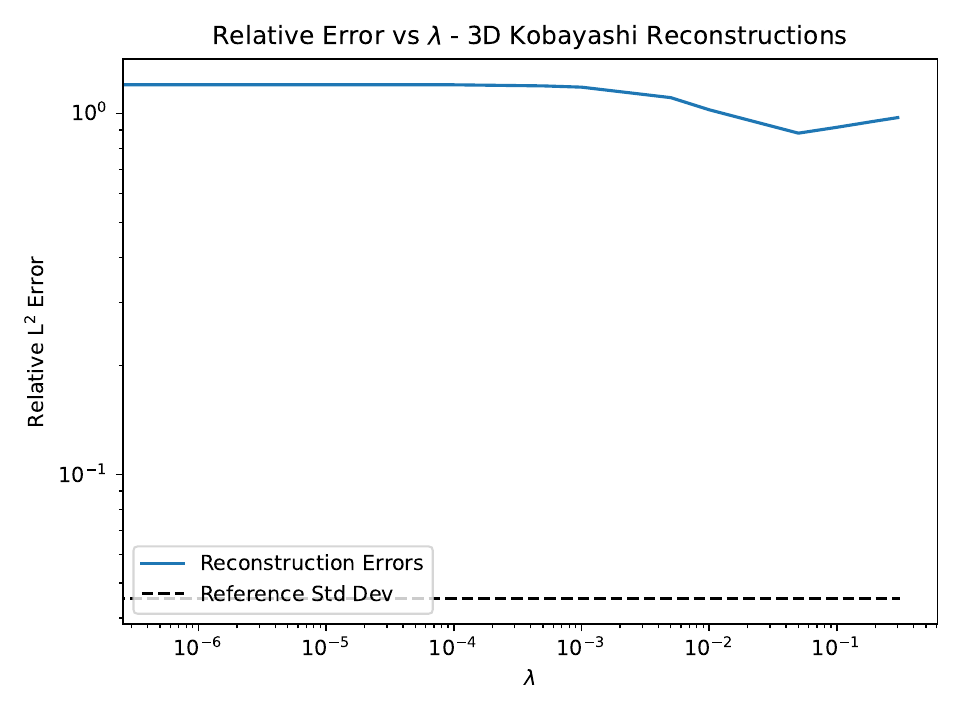}
    \includegraphics[width=0.65\linewidth]{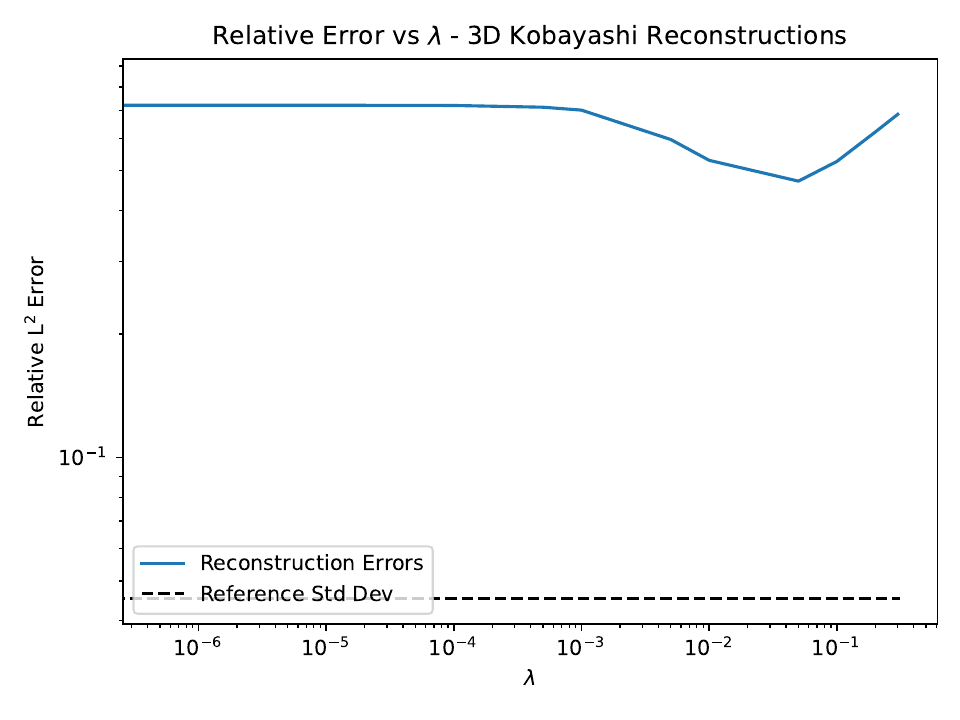}
    \includegraphics[width=0.65\linewidth]{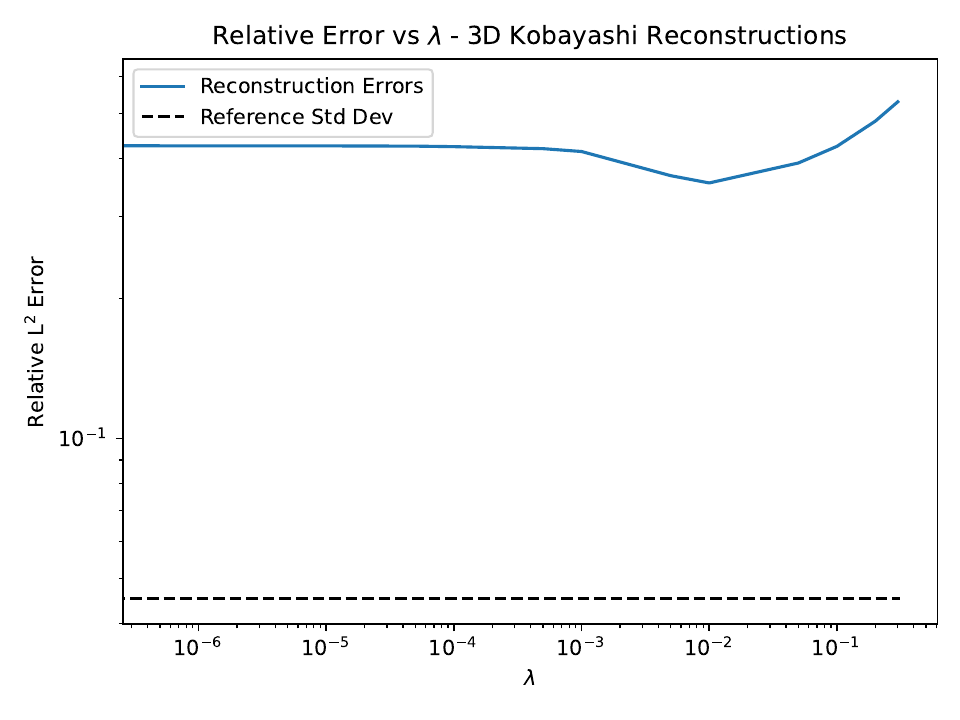}
    \caption{Errors associated with reconstructions of the Kobayashi problem. Top: 300 2$\times$2$\times$2 bins, 94.6\% memory reduction. Middle: 1000 2$\times$2$\times$2 bins, 82.22\% memory reduction. Bottom: 1500 2$\times$2$\times$2 bins, 73.3\% memory reduction.}
    \label{fig:3D kobayashi errors}
\end{figure}

The statistical errors associated with reconstructions of the Kobayashi problem are shown in Figure \ref{fig:3D kobayashi errors}.

\subsection{Modified Cooper-Larsen Problem Reconstructions Using Different Values of $\lambda$}
\subsubsection{Statistical Relative Errors in 2D Reconstructions}
\begin{figure}
    \centering
    \includegraphics[width=0.65\linewidth]{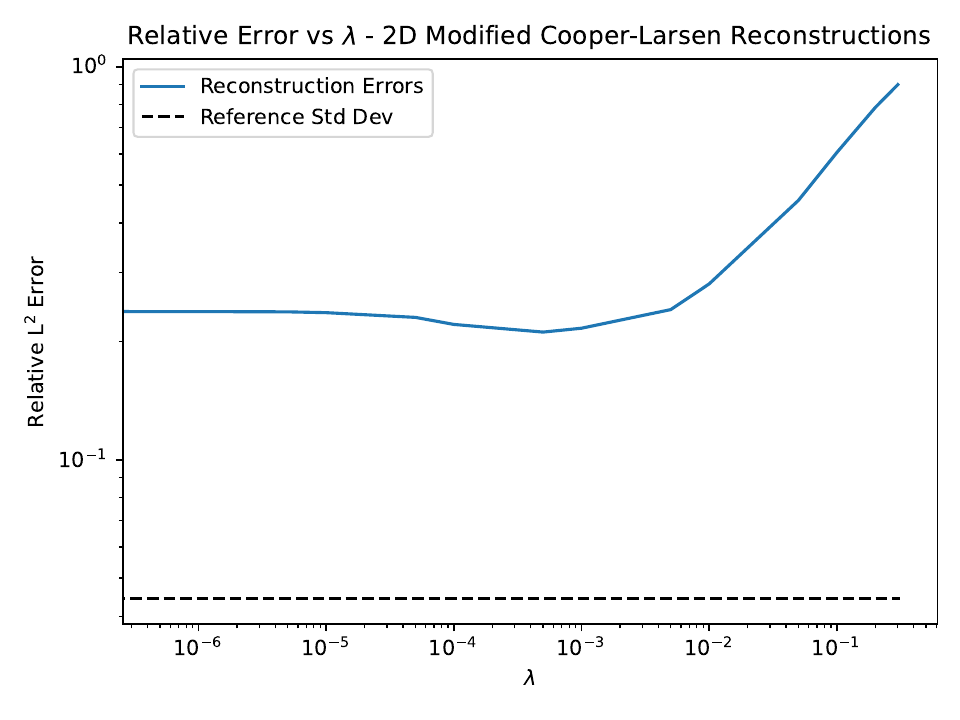}
    \includegraphics[width=0.65\linewidth]{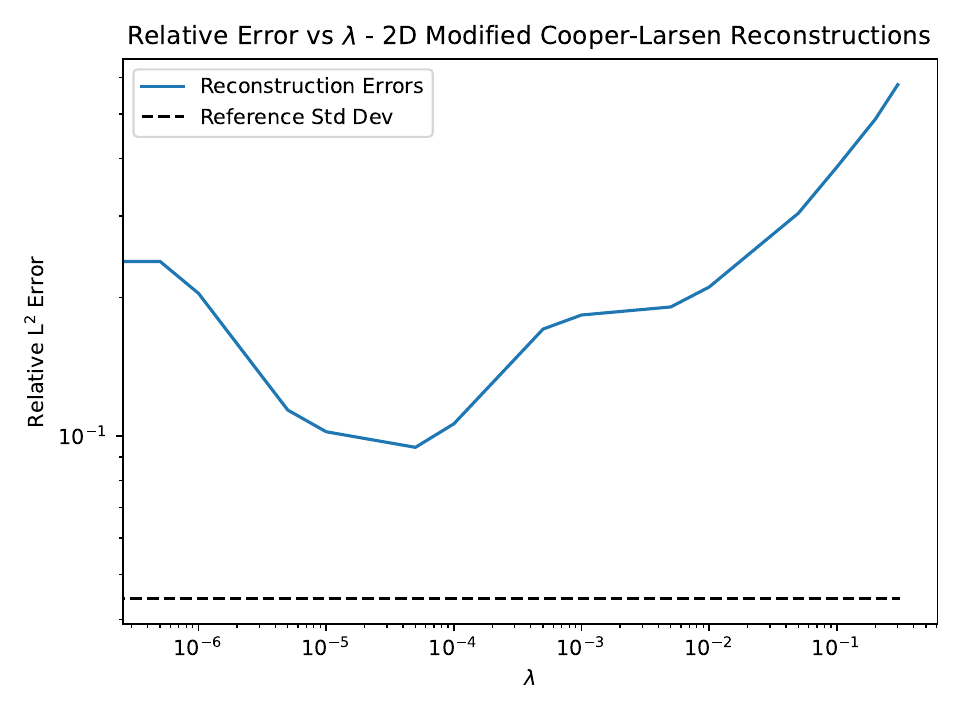}
    \includegraphics[width=0.65\linewidth]{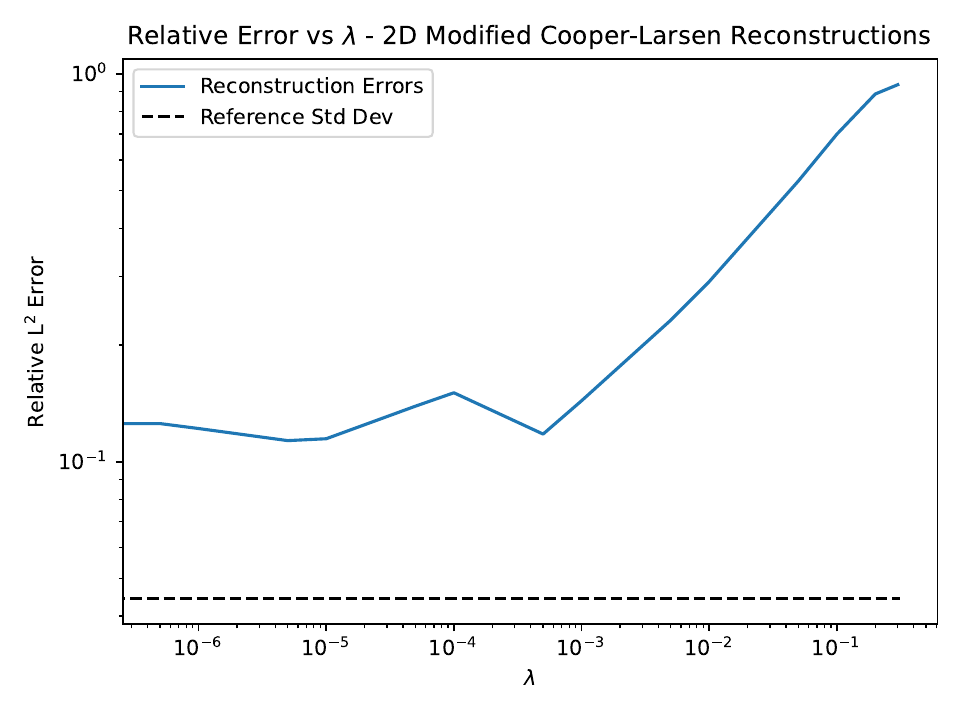}
    \caption{Errors associated with reconstructions of the modified Cooper-Larsen problem. Top: 300 3$\times$3 bins, 81.25\% memory reduction. Middle: 1000 3$\times$3 bins, 37.5\% memory reduction. Bottom: 1000 2$\times$2 bins, 37.5\% memory reduction.}
    \label{fig:2D cooper larsen errors}
\end{figure}

The statistical errors associated with reconstructions of the modified Cooper-Larsen problem are shown in Figure \ref{fig:2D cooper larsen errors}.

\subsection{Runtime}
\begin{figure}
    \centering
    \includegraphics[width=\linewidth]{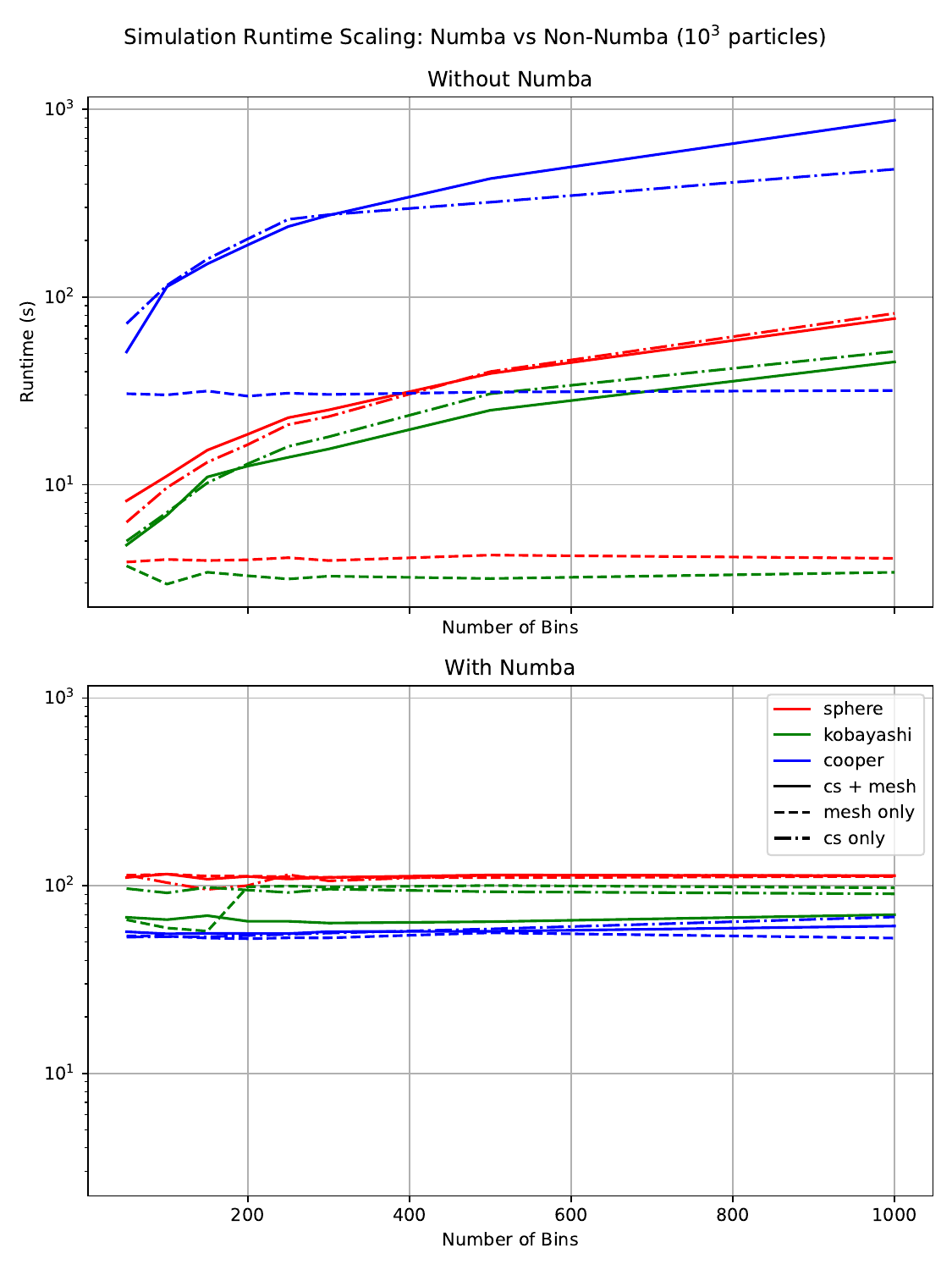}
    \caption{The runtime for MC/DC simulations with and without Numba for a variety of cases.}
    \label{fig:simulation runtime}
\end{figure}

As the number of coarse bins used in the problem increases, more calculations need to be performed for each particle, which increases the total simulation time. MC/DC can use Numba, a just-in-time compiler for Python, to achieve, in our case, an order of magnitude speedup. We can see the simulation runtime differences with and without Numba in Figure \ref{fig:simulation runtime}.

\begin{figure}
    \centering
    \includegraphics[width=0.63\linewidth]{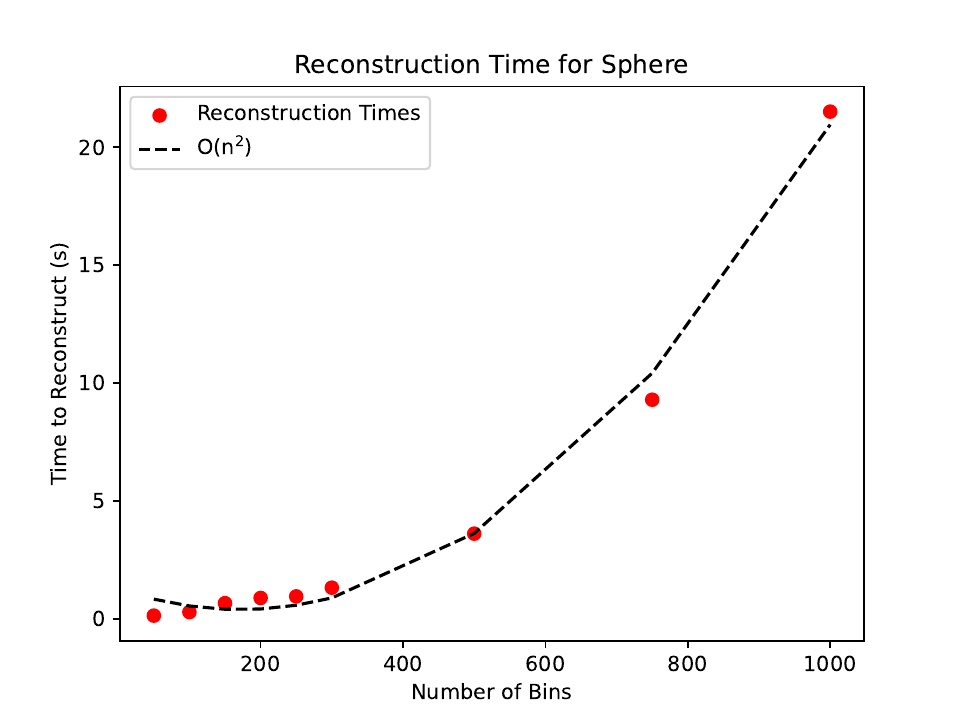}
    \includegraphics[width=0.63\linewidth]{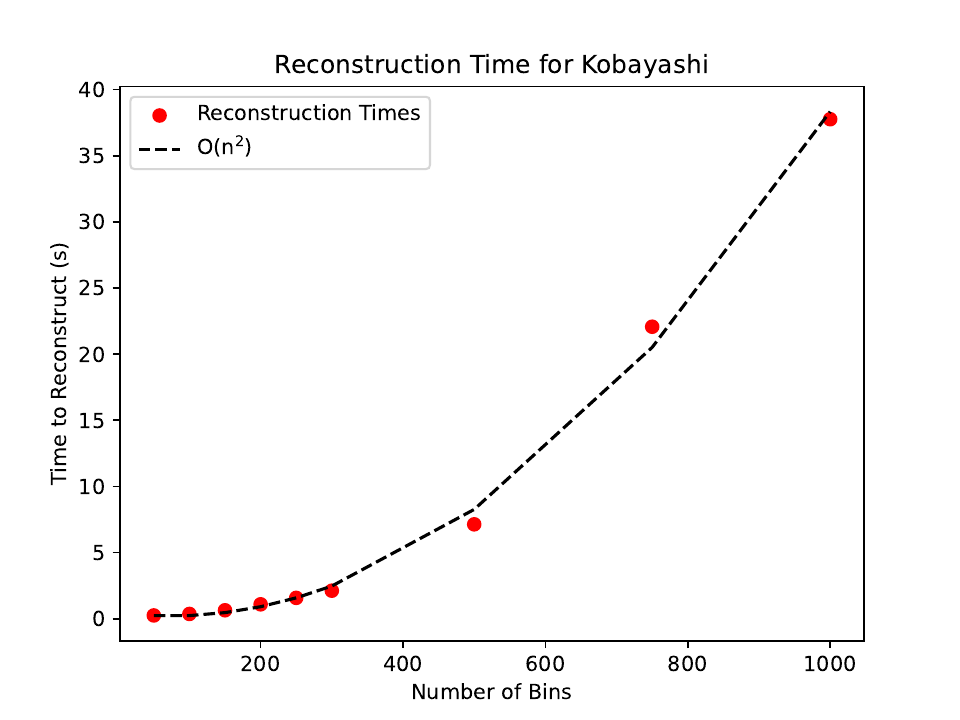}
    \includegraphics[width=0.63\linewidth]{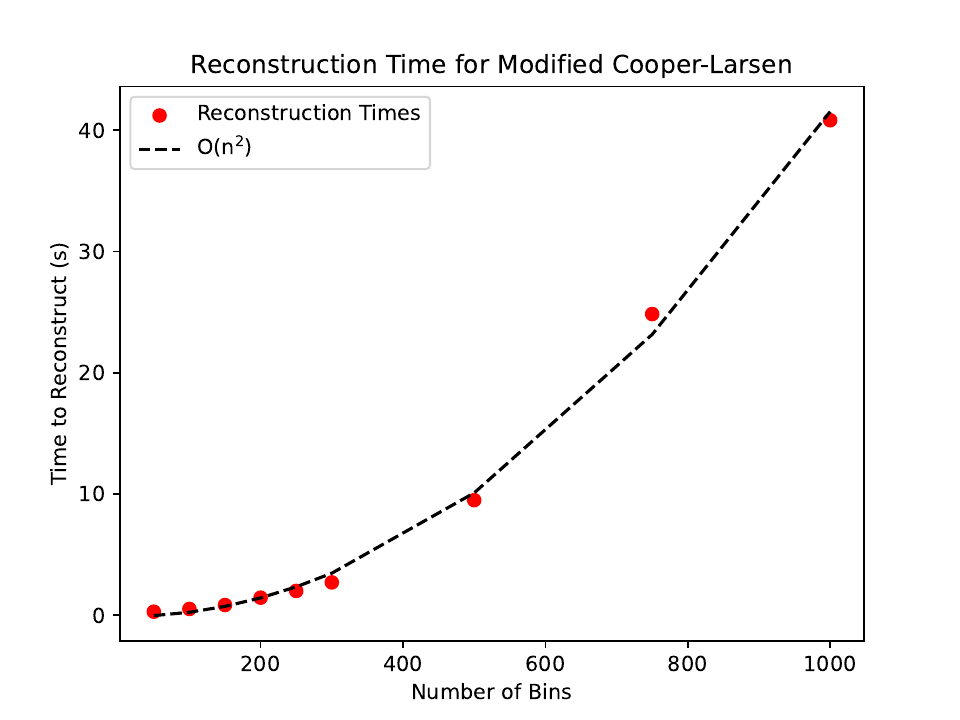}
    \caption{Time taken to find the reconstructions, as a function of the number of coarse bins. Fits for quadratic increase are plotted.}
    \label{fig:reconstruction runtime}
\end{figure}

The number of coarse bins used for tallying also changes the time taken for the basis pursuit denoising algorithm to find a solution in the sparse basis. With more coarse bins, there are more measurements to match, so the algorithm needs to search a larger space. The reconstruction runtimes as a function of the number of coarse bins are shown in Figure \ref{fig:reconstruction runtime}.

\newpage
\section{Discussion}
This method of compressed sensing - using overlapping coarse bins to collect tallies, and using basis pursuit denoising to find a reconstruction - reconstruct the given test cases, and it performs the best in cases with simple geometry. This is likely because the simple geometry is represented as a sparser vector in the DCT basis than the other geometries are.

Taking Figures \ref{fig:2D sphere recons} and \ref{fig:2D sphere errors} into consideration, we can see that in a simple 2D case with more than an 80\% reduction in memory usage, we are still able to achieve errors less than the standard deviation of the MC/DC solution. And, by increasing the number of bins in the same case, we can reduce the errors even further. Other cases do not fare quite so well, but still exhibit the trend of decreasing errors with increasing numbers of bins. 

The 3D sphere reconstructions (Figure \ref{fig:3D sphere errors}) show memory reductions of up to 96.25\% with associated errors within 2$\sigma$ of the reference solution mean. However, to achieve errors within 1$\sigma$ of the reference solution mean, memory reduction must be limited. There were no 3D reconstructions that achieved errors within 1$\sigma$ of the reference solution mean. It is possible that having even more bins would at some point push the errors below the 1$\sigma$ threshold, but it may take a prohibitively large number of measurements. For example, an 80\% reduction in memory for the sphere case (1600 bins) did result in smaller errors than the case with an 87.5\% reduction in memory (1000 bins), seen in Figure \ref{fig:3D sphere errors}.

As the number of coarse bins used increases, the accuracy of the reconstruction improves. However, the number of calculations performed in the simulation also increases. Figure \ref{fig:simulation runtime} shows that with Numba mode, the simulation runtime does not change based on the number of coarse cells. It does take more computational time for smaller numbers of coarse bins because Numba requires some overhead to compile the code at runtime. In the modified Cooper-Larsen problem, which contains a region with high scattering cross section, Numba provides an order of magnitude speedup with large numbers of bins. The trend in Figure \ref{fig:simulation runtime} also suggests that at larger and larger numbers of bins, the compilation overhead from Numba becomes less important, and its speedup capabilities become more useful.

An increasing number of coarse bins also results in a basis pursuit denoising optimization that is more difficult because there are more measurements to take into account. There are more degrees of freedom for the optimization to explore, so the search in the DCT basis takes more computational time as the number of measurements increases. It appears that as the number of coarse bins increases, the computation time to find a solution with basis pursuit denoising increases quadratically, as seen in Figure \ref{fig:reconstruction runtime}.

\section{Conclusions}
In this work, we introduce a method for compressed sensing to be applied to Monte Carlo neutron transport, by using large, overlapping tally bins to cover the problem space with fewer tallies than common Monte Carlo tallying would require. This methodology takes advantage of the fact that the global solution for neutron flux tallies - the quantity of interest in this work - can be represented as sparse in the discrete cosine transform basis. By searching for a sparse vector that matches the tallies in the coarse bins, we can find an approximation of the global neutron flux, while exhibiting large memory reductions, in some cases up to 96.5\%. Errors comparable to the reference solutions are achieved with 37.5\% memory reductions.

We investigate how the statistical relative errors of the reconstructions change as a function of the sparsity parameter $\lambda$. We find that there is some value of $\lambda$ that can provide the best reconstruction (i.e. the reconstruction with the smallest relative error), but that this value is different for different problems and for different resolutions for the same problem.

In 2 of the 14 cases presented here, we were able to successfully find reconstructions of the neutron scalar flux to within an accuracy of $1\sigma$ of the standard deviation of the reference solution, while also exhibiting substantial memory reductions (37.5\% and 81.25\%). However, this precision was only observed with the simplest geometry. For problems with more complex geometries, the solution can be reconstructed with an error that is within an order of magnitude of the standard deviation of the reference solution. Large ($>80\%$) memory reductions were observed in these cases.

The use of coarse tally bins to reduce memory in Monte Carlo neutron transport applications shows promise. Extending the coarse bin reconstruction approach to other dimensions (energy, angle, time) would likely reduce the memory requirements even further, though reconstruction times may reduce its applicability. Basis pursuit denoising was not chosen for its efficiency, and no parallelization was done, so the increased time taken for higher resolution reconstructions may be a limiting factor.

Other algorithms for searching the sparse basis for a solution are of interest, as the computational time taken to find a reconstruction appears to scale quadratically with the number of coarse bins used. A more quantitative investigation of the number of bins used and the trade-off between reconstruction accuracy and efficiency is necessary to assess whether gains in accuracy justify the additional computational cost.

Comparisons with a very high fidelity (very low statistical error and high-resolution tally grid) would be beneficial, and may be included in future publications as our software implementations are made to scale to larger high performance computing systems.

\section*{Disclosure statement}
The authors report there are no competing interests to declare.

\section*{Funding}
This work was supported by the Center for Exascale Monte-Carlo Neutron Transport (CEMeNT), a PSAAP-III project funded by the Department of Energy under Grant DE-NA003967.

\section*{Author Contributions}
C. Palmer, T. Palmer, and I. Variansyah conceived of the idea for the present work, helped develop methods, and provided feedback on this manuscript. E. Lame implemented the methods, performed all simulations and analysis, and wrote this manuscript.


\begin{thebibliography}{}
\bibitem[Candès, Romberg, and Tao(2005)]{StableSignalRecovery}
Candès, Emmanuel, Justin Romberg, and Terence Tao. 2005. ``Stable Signal
Recovery from Incomplete and Inaccurate Measurements.'' \emph{arXiv preprint}
math/0503066. https://arxiv.org/abs/math/0503066.

\bibitem[Donoho(2006)]{compressed_sensing}
Donoho, David. 2006. ``Compressed Sensing.'' \emph{IEEE Transactions on
Information Theory} 52 (5): 1289--1306.
https://doi.org/{10.1109/TIT.2006.871582}.

\bibitem[Candès and Romberg(2007)]{incoherence_constraint}
Candès, Emmanuel, and Justin Romberg. 2007. ``Sparsity and Incoherence in
Compressive Sampling.'' \emph{Inverse Problems} 23 (3): 969--985.
https://doi.org/{10.1088/0266-5611/23/3/008}.

\bibitem[Vaquer et al.(2016)]{vaquer2016}
Vaquer, Pablo~A., et al. 2016. ``A Compressed Sensing Framework for Monte Carlo
Transport Simulations Using Random Disjoint Tallies.'' \emph{Journal of
Computational and Theoretical Transport} 45 (3): 219--229.
https://doi.org/{10.1080/23324309.2016.1156550}.

\bibitem[Madsen(2017)]{madsen_2017_disjoint_tallies}
Madsen, Jonathan~Robert. 2017. ``Disjoint Tally Method: A Monte Carlo Scoring
Method Using Compressed Sensing to Reduce Statistical Noise and Memory.'' PhD
diss., Texas A\&M University.

\bibitem[Chen, Donoho, and Saunders(2001)]{basis_pursuit_denoising}
Chen, Scott~Shaobing, David~L. Donoho, and Michael~A. Saunders. 2001. ``Atomic
Decomposition by Basis Pursuit.'' \emph{SIAM Review} 43 (1): 129--159.
https://doi.org/{10.1137/S003614450037906X}.

\bibitem[Morgan et al.(2024)]{mcdc}
Morgan, Joanna, et al. 2024. ``Monte Carlo / Dynamic Code (MC/DC): An
Accelerated Python Package for Fully Transient Neutron Transport and Rapid
Methods Development.'' \emph{Journal of Open Source Software} 9 (94): 6415.
https://doi.org/{10.21105/joss.06415}.

\bibitem[Dalcín et al.(2005)]{mpi4py}
Dalcín, Lisandro, et al. 2005. ``MPI for Python.'' \emph{Journal of Parallel
and Distributed Computing} 65 (9): 1108--1115.
https://doi.org/{10.1016/j.jpdc.2005.03.010}.

\bibitem[Kobayashi et al.(2001)]{Kobayashi}
Kobayashi, Keisuke, et al. 2001. ``3D Radiation Transport Benchmark Problems and
Results for Simple Geometries with Void Region.'' \emph{Progress in Nuclear
Energy} 39 (2): 119--144. https://doi.org/{10.1016/S0149-1970(01)00007-5}.

\bibitem[Cooper and Larsen(2001)]{Cooper_Larsen}
Cooper, Marc~A., and Edward~W. Larsen. 2001. ``Automated Weight Windows for
Global Monte Carlo Particle Transport Calculations.'' \emph{Nuclear Science and
Engineering} 137 (1): 1--13. https://doi.org/{10.13182/NSE00-34}.

\bibitem[Owen(2017)]{Halton}
Owen, Art~B. 2017. ``A Randomized Halton Algorithm in R.'' \emph{arXiv preprint}
arXiv:1706.02808. https://arxiv.org/abs/1706.02808.

\bibitem[Diamond and Boyd(2016)]{CVXPY}
Diamond, Steven, and Stephen~Boyd. 2016. ``CVXPY: A Python-Embedded Modeling
Language for Convex Optimization.'' \emph{Journal of Machine Learning Research}
17 (83): 1--5.

\bibitem[Leppänen et al.(2025)]{serpent}
Leppänen, Jaakko, Ville Valtavirta, Antti Rintala, and Riku Tuominen. 2025. 
``Status of Serpent Monte Carlo Code in 2024.'' 
\emph{EPJ Nuclear Sci. Technol.} 11: 3. 
https://doi.org/{10.1051/epjn/2024031}.

\bibitem[Romano et al.(2015)]{openmc}
Romano, Paul K., Nicholas E. Horelik, Bryan R. Herman, Adam G. Nelson, Benoit Forget, and Kord Smith. 2015. 
``OpenMC: A State-of-the-Art Monte Carlo Code for Research and Development.'' 
\emph{Annals of Nuclear Energy} 82: 90--97. 
https://doi.org/{10.1016/j.anucene.2014.07.048}.

\bibitem[Kulesza et al.(2024)]{mcnp}
Kulesza, Joel A., et al. 2024. 
\emph{MCNP\textsuperscript{\textregistered} Code Version 6.3.1 Theory \& User Manual}. 
LA-UR-24-24602, Rev.~1. Los Alamos National Laboratory. 
https://doi.org/{10.2172/2372634}.


\end{thebibliography}
\end{document}